\documentclass[journal]{IEEEtran}

\usepackage{stmaryrd}
\usepackage{graphicx}
\usepackage{color}
\usepackage{url}
\usepackage{wrapfig}

\usepackage{tabularx}
\usepackage[tight]{subfigure}
\usepackage{balance}
\usepackage{comment}
\usepackage[justification=centering,small]{caption}

\usepackage{algpseudocode}
\usepackage{algorithm}
\usepackage{mathtools}

\newcommand{\superscript}[1]{\ensuremath{^{\textrm{#1}}}}

\def\wu{\superscript{*}}
\def\wg{\superscript{\dag}}
\def\wb{\superscript{\ddag}}
\def\wa{\superscript{\S}}

\renewcommand{\baselinestretch}{0.9}
\usepackage[left=.7in,right=.7in,top=.7in,bottom=.85in]{geometry}

\begin{document}

\title{Towards Physiology-Aware DASH:\\ Bandwidth-Compliant Prioritized Clinical Multimedia Communication in Ambulances}

\author{\IEEEauthorblockN{Mohammad Hosseini\wu, Yu Jiang\wb, Richard R. Berlin\wu\wg, Lui Sha\wu, Houbing Song\wa}
\IEEEauthorblockA{
  \begin{tabular}{cccc}
    \wu UIUC & \wb Tsinghua University & \wg Carle Foundation Hospital &\wa West Virginia University\\
  \end{tabular}
  ~\\
\wu \{shossen2, lrs\}@illinois.edu, \wb jy1989@mail.tsinghua.edu.cn, \wg richard.berlin@carle.com,\wa h.song@ieee.org}

\thanks{Mohammad Hosseini and Lui Sha are with the Department of Computer Science, University of Illinois at Urbana-Champaign (UIUC), Illinois, USA.}
\thanks{Yu Jiang is with the School of Software, Tsinghua University, China.}
\thanks{Richard Berlin is with the Department of Computer Science, UIUC, and Carle Foundation Hospital, Urbana, Illinois, USA.}
\thanks{Houbing Song is currently with the Department of Electrical, Computer, Software, and Systems Engineering, Embry-Riddle Aeronautical University, Daytona Beach, USA.}
\thanks{Correspondence authors: Mohammad Hosseini (shossen2@illinois.edu) and Yu Jiang (jy1989@mail.tsinghua.edu.cn).}}

\maketitle

\begin{abstract}
The ultimate objective of medical cyber-physical systems is to enhance the safety and effectiveness of patient care. To ensure safe and effective care during emergency patient transfer from rural areas to center tertiary hospitals, reliable and real-time communication is essential. Unfortunately, real-time monitoring of patients involves transmission of various clinical multimedia data including videos, medical images, and vital signs, which requires use of mobile network with high-fidelity communication bandwidth. However, the wireless networks along the roads in rural areas range from 4G to 2G to low speed satellite links, which poses a significant challenge to transmit critical patient information.

In this paper, we present a bandwidth-compliant criticality-aware system for transmission of massive clinical multimedia data adaptive to varying bandwidths during patient transport. Model-based clinical automata are used to determine the criticality of clinical multimedia data. We borrow concepts from DASH, and propose physiology-aware adaptation techniques to transmit more critical clinical data with higher fidelity in response to changes in disease, clinical states, and bandwidth condition. In collaboration with Carle's ambulance service center, we develop a bandwidth profiler, and use it as proof of concept to support our experiments. Our preliminary evaluation results show that our solutions ensure that most critical patient's clinical data are communicated with higher fidelity.
\end{abstract}
\IEEEpeerreviewmaketitle

\section*{Acknowledgment}
This paper reviews a technology package that is the result of a team effort. Lui Sha led the development of model-driven multimedia clinical systems and all system integration. Yu Jiang and Houbing Song helped with the development of physiological and cyber-physical models. Richard Berlin guided the study of patient emergency care, physiological multimedia data, and medical correctness. Mohammad Hosseini led the design and development of the multimedia communication system, adaptation system, heuristic algorithms, as well as development of the profiler app and conducting experiments and evaluations.

The material presented in this paper is based upon work supported in part by NSF CNS 1329886, by NSF CNS 1545002, and in part by ONR N00014-14-1-0717. Any opinions, findings, and conclusions or recommendations expressed in this publication are those of the authors and do not necessarily reflect the views of the grant providers.

\section{Introduction}
\label{sec:introduction}
There is a great divide in emergency medical care between rural and urban areas. The highest death rates are found in rural counties, which has motivated huge research efforts in recent years to enhance the safety and effectiveness of patient care especially in these areas. For mobile care and emergency ambulance patient transport from rural areas to center tertiary hospitals, time to definitive treatment is critical. During emergency rural patient transport, the real-time monitoring of a patient by the physicians at the receiving regional center hospital provides vital assistance to the Emergency Medical Technicians (EMT) in the ambulance. Such simultaneous monitoring allows the physicians in the center hospital to remotely supervise the patient in the ambulance and to help follow best treatment practices based on patient's condition and clinical multimedia data. These multimedia data are often generated from various clinical sensors, and include clinical videos, medical images, speech data, voice communication, and vital signs which overall form a rich clinical multimedia system. These clinical multimedia data are then transmitted and continuously monitored at the center hospital for diagnosis, best-practice orders and treatments.

\begin{figure}[!t]
\centering
\includegraphics[width=.95\columnwidth]{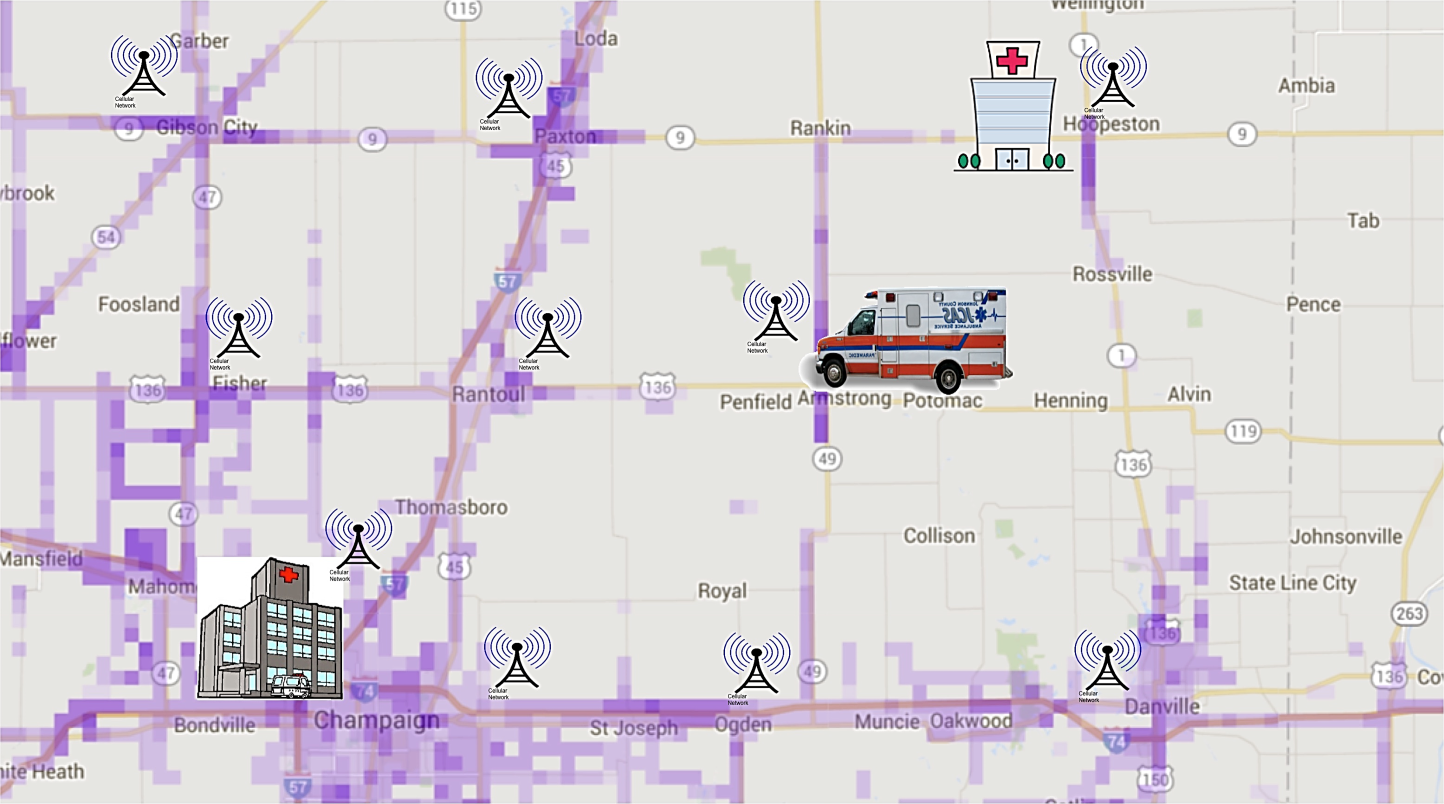}
\caption{Real-world mobile data coverage maps between Hoopeston rural hospital and Carle regional center hospital (Sprint carrier). Darker colors show higher bandwidth.}
\label{fig:coverage}
\end{figure}
To ensure safety and effectiveness of emergency care during such ambulance patient transfer, reliable and real-time communication is essential. However, the wireless networks along the roads in rural areas range from 4G to 2G to low speed satellite links, which can result in periods of low fidelity communication bandwidth. Figure \ref{fig:coverage} illustrates a real-world example. It demonstrates the 4G mobile data coverage maps along possible routes from a rural hospital located in Hoopeston, Illinois, to the Carle \cite{carle} regional center hospital in Urbana, Ilinois; the darker colors show higher bandwidth. As the map demonstrates, the roads located in the rural areas manifest varying and often poor communication coverage especially when compared to coverage in urban areas. Therefore the challenges for transmission of necessary clinical multimedia data may be exaggerated during high-speed ambulance patient transport. The criticality of clinical data needed for remote supervision also varies as the patient condition may change. As a result, the semantics relation between limited and dynamically-changing communication bandwidth and the physiological criticality of communicated clinical multimedia becomes a vital problem which unfortunately remains unaddressed by prior work.

In this paper, we propose a bandwidth-compliant criticality-aware system for adaptive transmission of clinical multimedia data in response to varying bandwidths during ambulance patient transport. We use the concepts of model-driven clinical automata to differentiate the criticality of each category of clinical multimedia data at any given clinical state. We take initial steps towards extending DASH towards physiology awareness, and present adaptation techniques to transmit more critical multimedia data with higher fidelity as the disease and clinical states change. Our approach is to selectively choose appropriate clinical multimedia data sampling frequencies given their physiological criticality so that the bandwidth requirements of clinical multimedia streams best satisfies a limited bandwidth available to an ambulance. In collaboration with Carle's ambulance service center, we develop a bandwidth profiler, and use it as proof of concept to support our experiments. Our preliminary evaluation results show that our physiology-aware bandwidth-compliant adaptations ensure that most critical patient's clinical multimedia data are communicated with higher fidelity.

The paper is organized as follows. In Section \ref{sec:background}, we cover a wide area of background and related work, and discuss how our system is related to them. In Section \ref{sec:scenarios}, we discuss real clinical use-cases of prioritization in emergency acute disease care. Section \ref{sec:methodology} explains our methodology for adaptive clinical multimedia data management, including the problem definition and the adaptation algorithms. Our evaluation results are presented in Section \ref{sec:evaluation}, while in Section \ref{sec:conclusion} we conclude the paper and briefly discuss possible avenues of future work.
\section{Background and Related Work}
\label{sec:background}
In this section we present different concepts and categories of work relevant to our proposed system.

\subsection{Dynamic Adaptive Streaming}
One of the main approaches for bandwidth saving when bandwidth is limited and dynamically changes is adaptive streaming. Adaptive streaming in general is a process where the quality of a data stream is altered in real-time while it is being transmitted. The adaptation of quality is controlled by decision modules on either the client or the server, and can be the result of adjusting various network or device metrics.
For example, with a decrease in network throughput, adaptation to a lower bitrate may reduce packet loss and therefore improve the user's experience.


Researches on adaptive streaming have been mostly applied to video context, and generally range from network coding, bandwidth detection, and rate determinations \cite{cmm1,cmm2,mahbub1,xiao1,xiao2} to quality of service and user experience \cite{cmm3,mao1,mao2}; many also have been applied to video-based tele-medicine systems \cite{telemedicine1,telemedicine2,telemedicine3,cicalo,rehman1,rehman2}. In \cite{cicalo}, \cite{rehman1}, and \cite{rehman2} for instance, the authors study bandwidth adaptation techniques for adaptive transmission of medical videos over the bandwidth-limited mobile networks, some targeting 4G and beyond small cells. Also, the authors in \cite{qiao} target lower network layers in mobile tele-medicine, and propose a TDMA-based MAC layer scheme for bandwidth reservation among cells for tele-medicine traffic handoff.

Dynamic Adaptive Streaming over HTTP (DASH) specifically, also known as MPEG-DASH \cite{dash1}, is probably the most prominent example which has been applied in the context of video streaming. It is an ISO standard that enables adaptive bitrate streaming whereby a client chooses the video segment with the appropriate bitrate based on its constrained resources such as bandwidth. Some work studied feasibility of extending DASH into domains other than videos and different variations, such as adaptive 3D graphics streaming \cite{hosseini1, hosseini3}, energy-aware DASH \cite{movid15, edash1}, and viewport-aware DASH for adaptive streaming of Virtual Reality (VR) contents \cite{ism16}. 
In this paper, we extend similar concepts towards clinical domain, and run an initial study towards a physiology-aware DASH. We apply adaptive streaming to the context of clinical multimedia transmission in emergency patient transport scenarios where a high-speed ambulance en route from a rural hospital to a center hospital is encountered with limited and variable bandwidth.

\subsection{Prioritized and Criticality-Aware Adaptations}
Generally the priority of contents being transmitted in a network can have different importance or criticality given the context or various settings. A large body of work have proposed prioritized approaches for a differentiated transmission service among contents of various priorities. For example in \cite{hosseini1}, and \cite{hosseini2} 
the authors adopted prioritization techniques towards the study of efficiently transmitting 3D contents to resource-limited devices given the importance of various objects in the gaming context. The authors in \cite{priority1} applied the concept of prioritization on a multi-user teleconference room equipped with a static camera capturing the whole room. 
Similarly, some works in the context of large-scale immersive systems \cite{mmve} and \cite{mmsj} studied data prioritization in regards to bandwidth savings. 
In another work \cite{criticality1}, the authors used similar concepts, and proposed a bitrate allocation approach to minimize energy consumption in wireless surveillance systems according to event criticality. The authors in \cite{criticality2} proposed criticality-aware access control policies to control resource access for both critical and non-critical events within smart spaces. They add a new criticality parameter to measure the urgency of a critical event. Similarly, in \cite{criticality3}, the authors propose a criticality-aware clustering protocol in the context of wireless sensor networks. In their approach, they keep a criticality threshold for information, and if such information is sensed, it is sent with highest priority to the base station.
While all these works employ the concept of prioritization according to contents' criticality, the application of clinical data prioritization and criticality awareness in the context of clinical and medical domain has remained unaddressed by the existing work.

The most related work to our system is \cite{criticality4}, in which the authors use the notion of prioritization and criticality awareness for pervasive medical sensor networks. Their model enables different types of access control decisions and assigns highest priority to critical operations and lowest priority to normal access control situations. While this work uses criticality awareness in the context of medical domain, their notion of prioritization is only tied to access control decisions. However, the relation between clinical multimedia transmission, bandwidth consumption, and criticality of multimedia data in various physiological states has not been addressed heretoforth.

\subsection{Clinical Optimization}
Most existing optimization approaches developed in the clinical domain target high-level algorithmic clinical practices, and provide optimization in clinical workflow, services, and emergency care management \cite{hosseini2015sink,wu2015safe}. For emergency care for instance, optimized medical best practices have been created for patients in major hospitals, such as \cite{CMA}, \cite{european2008guidelines}, and \cite{party2008national}. 

The University of Texas' MD Anderson Cancer Center for example has developed clinical management algorithms \cite{CMA} that provides a high-level workflow optimization for diagnosis, evaluation, and treatment of specific acute diseases. G\"{o}rlitz \textit{et. al.} also studied the feasibility of an optimized stroke manager service concept using a combined service and software engineering approach, and improved workflow and IT architecture for enhanced post-stroke management~\cite{Gorlitz2012HPT}. Hofmann et al described concepts used for process optimization in acute disease care, and evaluated industrial methods to provide quality improvement in emergency care management \cite{Hofmann2012PM}. Panzarasa and Stefanelli likewise designed an optimized evidence-based workflow management system as components of a knowledge management infrastructure by efficiently exploiting the available knowledge resources, aiming to maximize the performance of clinical services~\cite{Panzarasa2006NS}.

While the proposed concepts in these studies consider requirements for clinical optimizations, they are mainly designed for well-equipped major medical centers around high-level clinical services, workflow, and emergency care management. Many fundamental disease-specific system and pathophysiological issues, especially problems associated with communication bandwidth, mobility, and criticality of clinical data still remain unaddressed.

\subsection{Executable Clinical Automata}
From a medical perspective, physicians are taught organ system function as part of the representation of disease process. They look for patterns of pathophysiological changes (the change in physiological measurements as a result of disease) within an organ system to understand organ states. 
This organ-centric view of pathophysiological expression also matches medical treatment, which is captured by best-practice medical workflows. Therefore, our adaptation system is plugged into, and fed with model-driven executable medical automata including executable best-practice workflow models, and model-based clinical automata such as disease and organ system automata \cite{hosseini2016pathophysiological}.


\begin{figure}[!htbp]
\centering
\subfigure[Rural Hospital]{
    \centering
    \includegraphics[width=1\columnwidth]{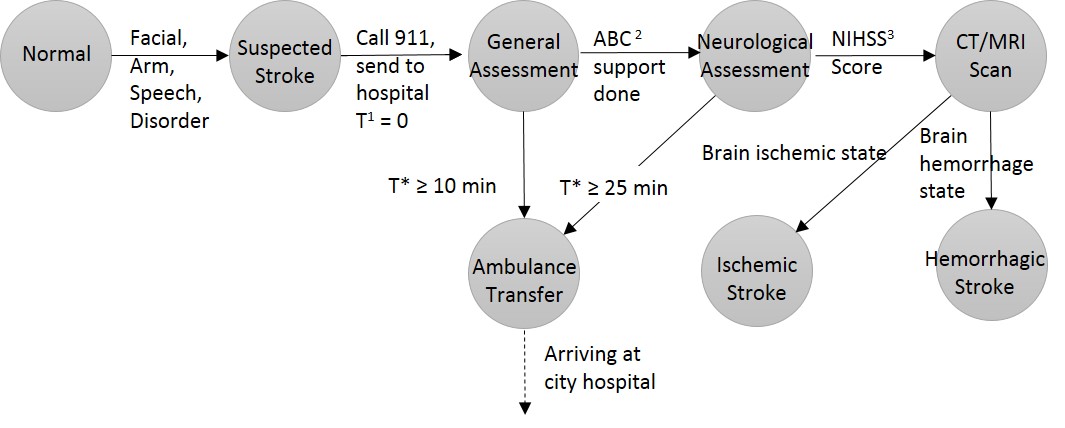}
	\label{fig:strokeRuralAutomata}}
\subfigure[Ambulance]{
    \centering
    \includegraphics[width=.9\columnwidth]{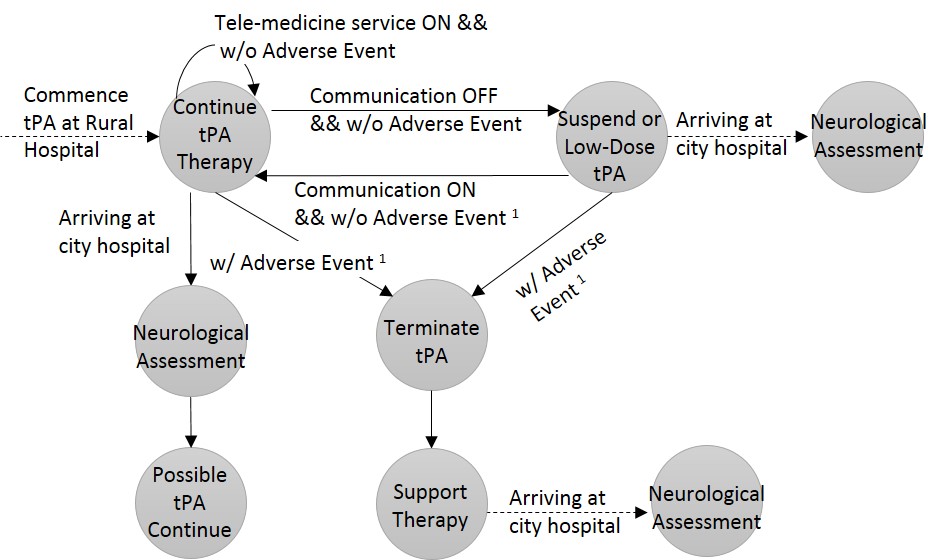}
	\label{fig:strokeAmbulanceAutomata}}
\subfigure[Center Hospital]{
	\centering
	\includegraphics[width=\columnwidth]{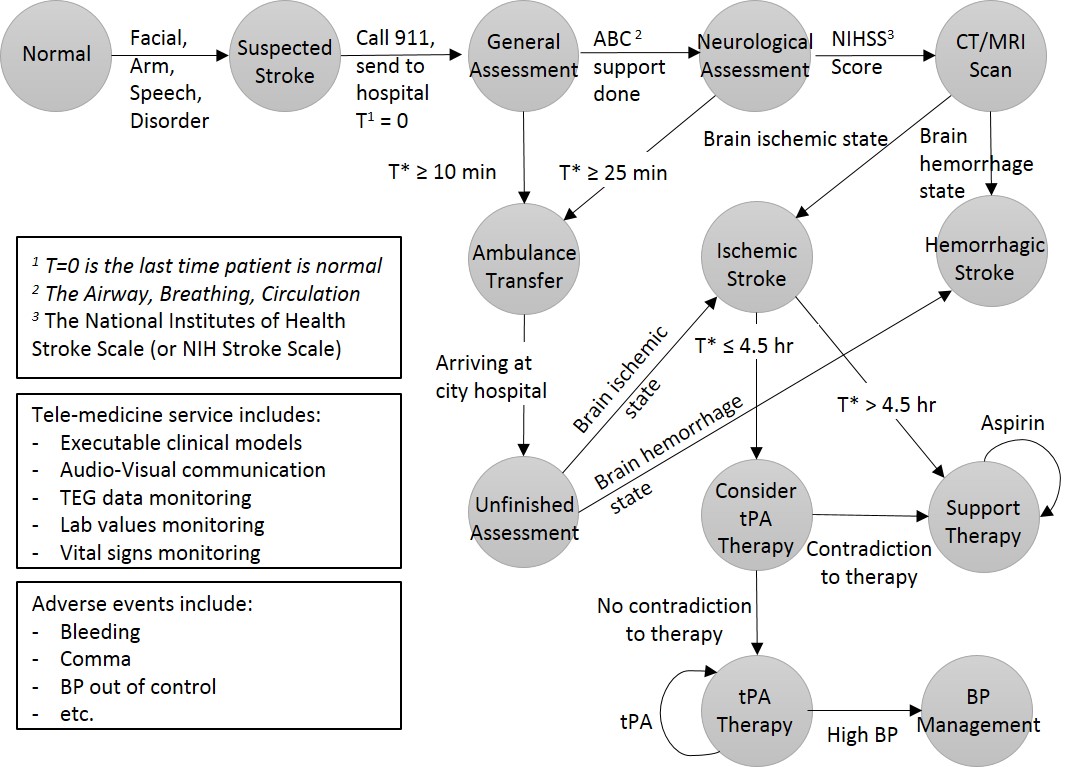}
	\label{fig:strokeCenterAutomata}}
	
\caption{An instance of simplified stroke best-practice automata distributed between a rural hospital, ambulance, and a center hospital.}
\vspace{-.5cm}
\label{fig:strokeAutomata}
\end{figure}

\begin{figure*}[!t]
\centering
\hspace{-.5cm}
\includegraphics[width=1.6\columnwidth]{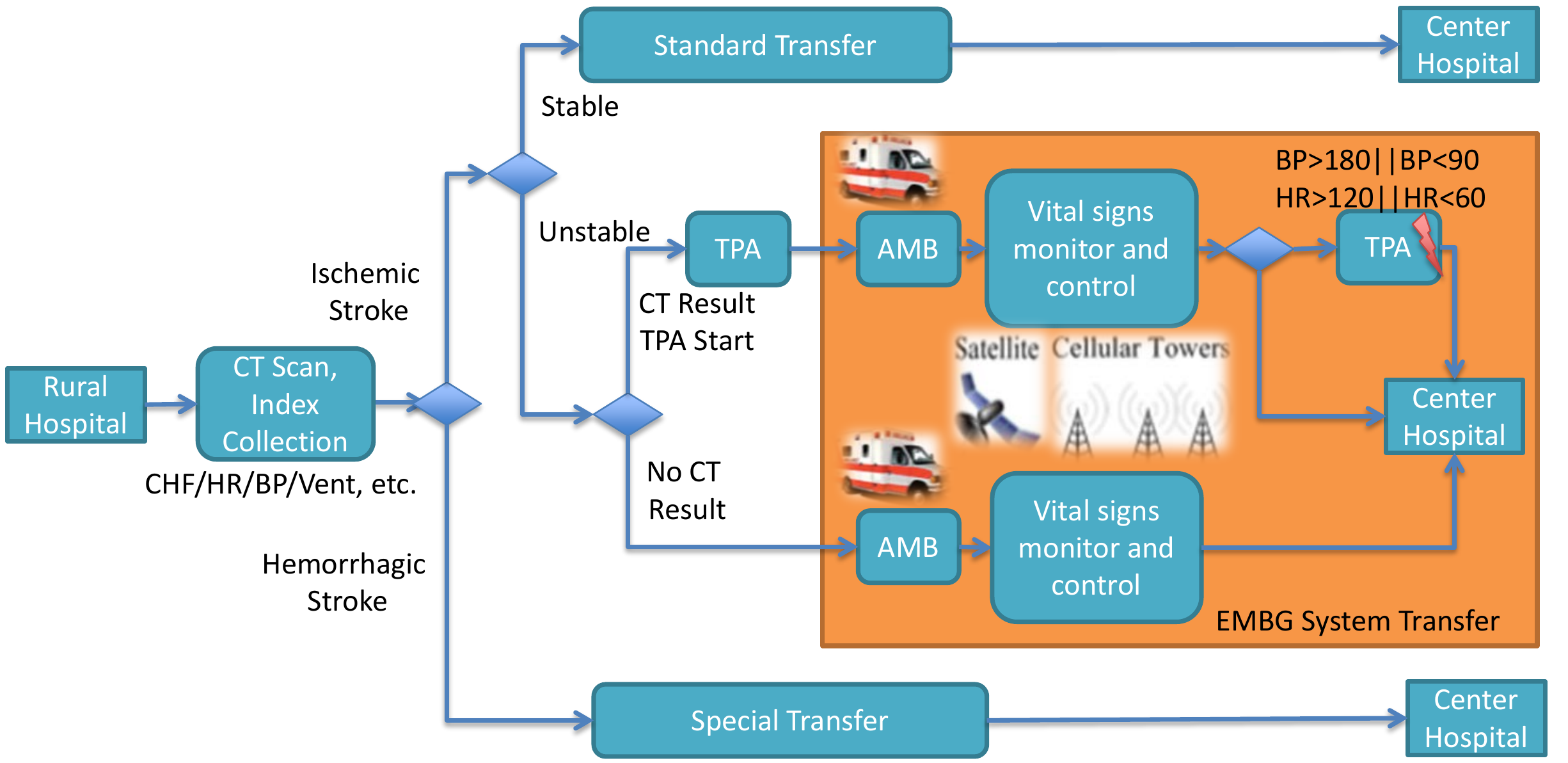}
\caption{Envisioned workflow for stroke patient care from a rural hospital to a regional center hospital.}
\label{fig:example-tpa}
\vspace{-.5cm}
\end{figure*}

Figure \ref{fig:strokeAutomata} shows three greatly simplified best-practice automata for stroke that are executed at a rural hospital (Figure \ref{fig:strokeRuralAutomata}), an ambulance (Figure \ref{fig:strokeAmbulanceAutomata}), and at a supervising regional center hospital (Figure \ref{fig:strokeCenterAutomata}). The states in each automaton represents different physiological states of an organ. The physiological changes of patients which result in satisfaction of the condition for a new organ state causes state transitions within these automata. The state changes can also be driven when the physicians confirm the new patient and organ states. Possible variances in capabilities, expertise, and physical environment can cause different levels of abstraction for executable clinical automata and different sets of generated clinical multimedia data at each location. The best-practice models are executed in real-time, allowing doctors at center hospitals to retrieve necessary clinical multimedia information and supervise a patient remotely by sending appropriate commands. The engine of our system in this work is models of executable clinical automata, and we use the concepts of executable automata to model all clinical components including patient and disease automata. We extend the same concepts towards communication, and codify bandwidth available to a high-speed ambulance transferring a patient from a rural hospital to a center hospital as an executable automata. A model-driven communication automata also allows expert physicians to remotely check and track the communication when necessary. Our system therefore semantically links the communication bandwidth to other model-driven clinical components including best-practice workflow automata, disease automata, and organ automata. It therefore integrates all executable clinical models together and provides model-driven coordination and synchronization among various clinical components.

In summary, in this paper we build upon concepts from the aforementioned related work, and propose bandwidth-compliant prioritized adaptations for efficient delivery of various clinical multimedia for emergency care. Our adaptation is applied to the context of emergency patient transport of acute diseases within a bandwidth-limited high-speed ambulance in regards to the criticality of clinical multimedia data given the patient's physiological states. The design of our bandwidth-compliant clinical multimedia adaptation system takes into consideration the variability of priorities within different clinical components that may impact the emergency care based on interaction among various models of clinical components. The key components include patient physiological models including disease and organ models, patient condition models, and models of communication bandwidth. Our proposed system aims to enhance the effectiveness of emergency patient care from a high-speed ambulance leveraging an efficient interaction among all clinical models, which all change throughout the transport.
\section{Prioritization in Emergency Patient Care}
\label{sec:scenarios}
Various clinical multimedia data can have different priorities depending on the context of specific episodes of acute patient care transport. We describe acute stroke scenario as a real-world use case to illustrate the concept.
Figure \ref{fig:example-tpa} illustrates the workflow for a stroke patient being transferred from a rural facility to a regional hospital center. Let's consider a 70 year old male patient arrives at a rural hospital and the diagnosis of acute stroke is considered. Initially a CT (Computerized Tomography) head scan is performed. The CT images are sent electronically to the regional center for interpretation. At this clinical state, highest priority is devoted to transmission of streams associated with CT images. The patient's neurological examination, laboratory data, and vital signs (including heart rate (HR), blood pressure (BP), oxygenation level) are obtained and sent for the purpose of continuous monitoring. The diagnosis of an acute stroke is made, and the patient is placed in an ambulance for transport. Physicians at the center hospital remotely monitor the physiological models including patient models and disease models. Models of clinical best practices and communication bandwidth models are also executed continuously in interaction with each other.

A video camera and microphone in the ambulance, connected to the regional center, is used to remotely monitor the patient's status through audio-visual data during ambulance transport. In accordance with the patient's physiological models, it is determined that the patient has a hemorrhagic stroke (bleeding into the brain from blood vessel rupture). In this case, models of clinical best practices suggest that time of transport is most important. Available bandwidth is used for communication with specialists at the regional center in case emergency consultation, interventional radiology or surgery is indicated. The HR, BP, oxygenation, and neurological status are remotely monitored. In these situations, with higher bandwidth according to the communication bandwidth models, the audio/video support and therefore, the transmission of audio-visual streams gets very important. However, in the event of limited bandwidth, the audio-visual monitoring system as well as the transmission of repeated laboratory data get limited with secondary priority. Highest priority in this situation is maintaining the patient's vital signs. The HR and BP in specific must be kept within strict limits. The BP assumes particular importance if it rises too high (greater than 180/-) or falls too low (less than 90/-) which are indicated by the physiological models. In accordance with guidance given by best-practice clinical models, audio communication with the center hospital to manage elevated blood pressure assumes highest priority if BP is greater than 180. This might require the continuous intravenous infusion of active medications to lower blood pressure in the ambulance, using nicardipine or nitroprusside medications. Once communication models show availability of higher bandwidth, periodic laboratory data, treatments for an elevated blood glucose (greater than 350), and the video camera streams can be used with normal sampling frequency of transmission with higher priority levels.

\begin{figure*}[!t]
\centering
\includegraphics[width=1\textwidth, trim = 50 60 30 60, clip = true]{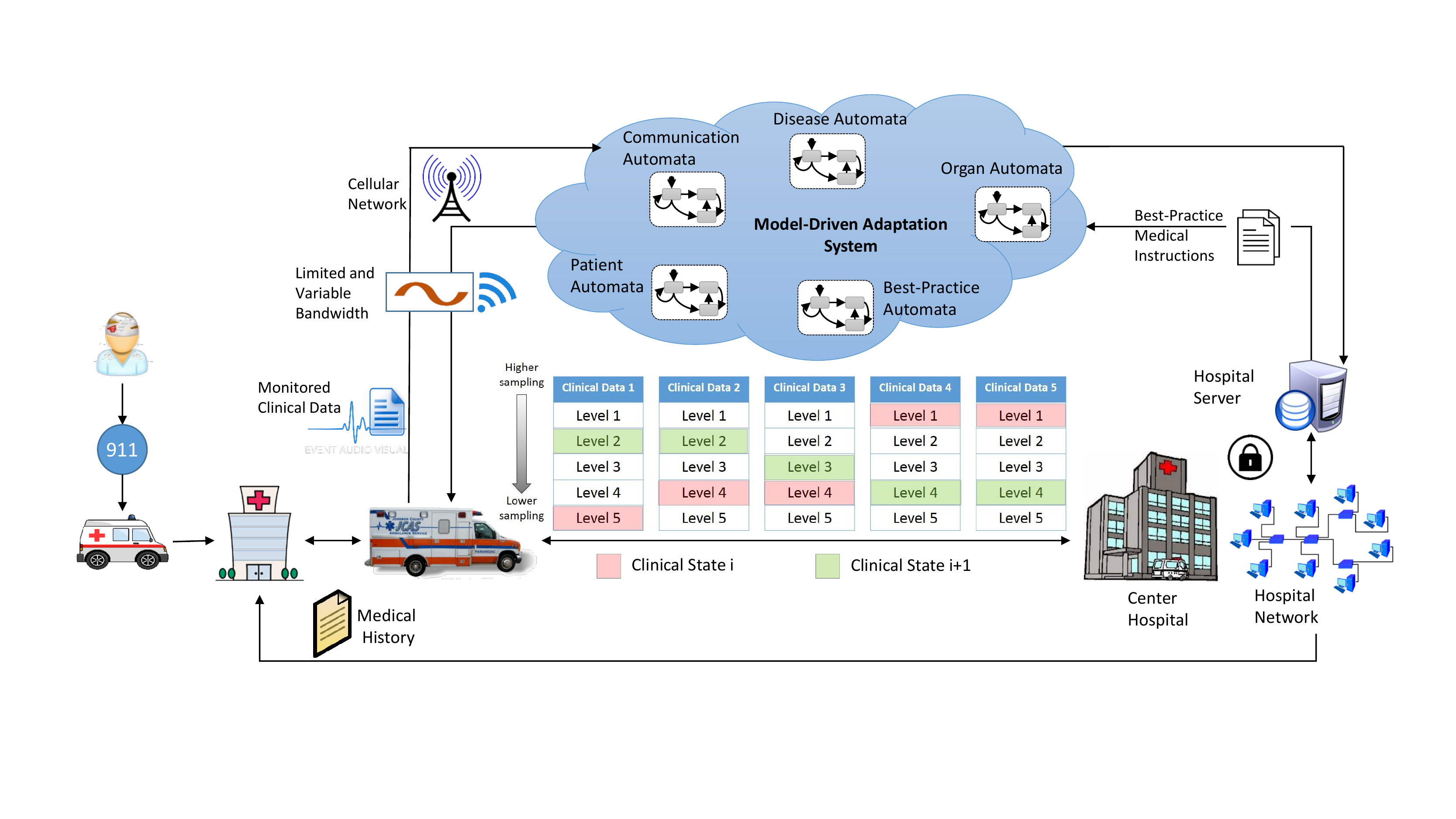}
\vspace{-1cm}
\caption{An example scheme of proposed adaptation approach.}
\vspace{-.5cm}
\label{scheme}
\end{figure*}

Now for a different patient, let's consider a 70 year old male patient who has an ischemic stroke (a clotted artery reducing blood flow to the brain rather than hemorrhage into the brain itself). In this case, blood coagulation laboratory data, TEG\footnote{Thromboelastography (TEG) is a method of testing the efficiency of blood coagulation, which helps with timing clot dissolving therapy.} examination, and the early transport treatment with a clot dissolving medication such as tissue plasminogen activator (tPA) infusion may be recommended given inputs from the best-practice clinical models. The patient is placed in the ambulance, tPA is continuously running, and ambulance transport towards the city hospital follows. In the ambulance, if the patient has stable vital signs based on the current state of physiological models, then priority attention is turned to a) the management of the tPA infusion, b) the maintenance of BP within strict parameters, and c) voice communication for the treatment of complicating factors. The ambulance crew and regional center physicians work as a team managing the patient's physiological status and tPA infusion. In this clinical state, the transmission of BP data, as well as voice communication has higher priority compared to other clinical data. If there is a state change in the patient's physiological models leading to signs of deterioration in the patient's condition, a complication of the tPA or extension of the area of brain with limited blood flow is assumed. Therefore, here bandwidth attention turns to managing the patient's vital signs, while the clinical data associated with tPA itself, coagulation values, TEG examination data, and laboratory results are considered with secondary importance. Once patient neurological deterioration is determined on the physiological models, priority is directed to the vital signs and maintaining oxygenation. Therefore, intubation and artificial ventilation are considered higher priority, while the laboratory data such as hematocrit (percentage amount of red blood cells in the blood) have lower priority given the available bandwidth.

\section{Proposed System}
\label{sec:methodology}

\begin{figure}[!htbp]
\centering
\includegraphics[width=.8\columnwidth]{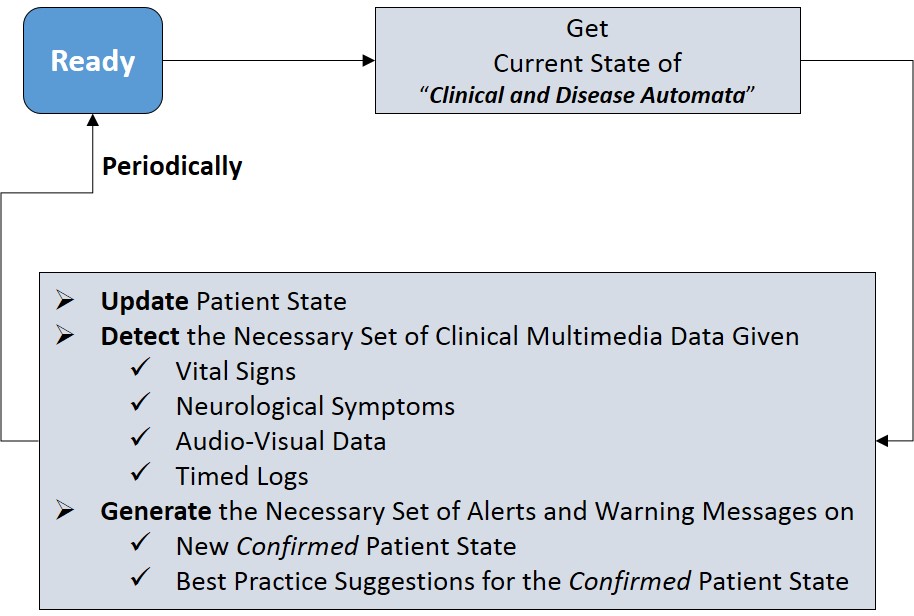}
\caption{Emergency care manager abstract.}
\label{manager}
\end{figure}

\begin{figure*}[!htbp]
\centering
\includegraphics[width=1.6\columnwidth, trim = 108 10 205 20, clip = true]{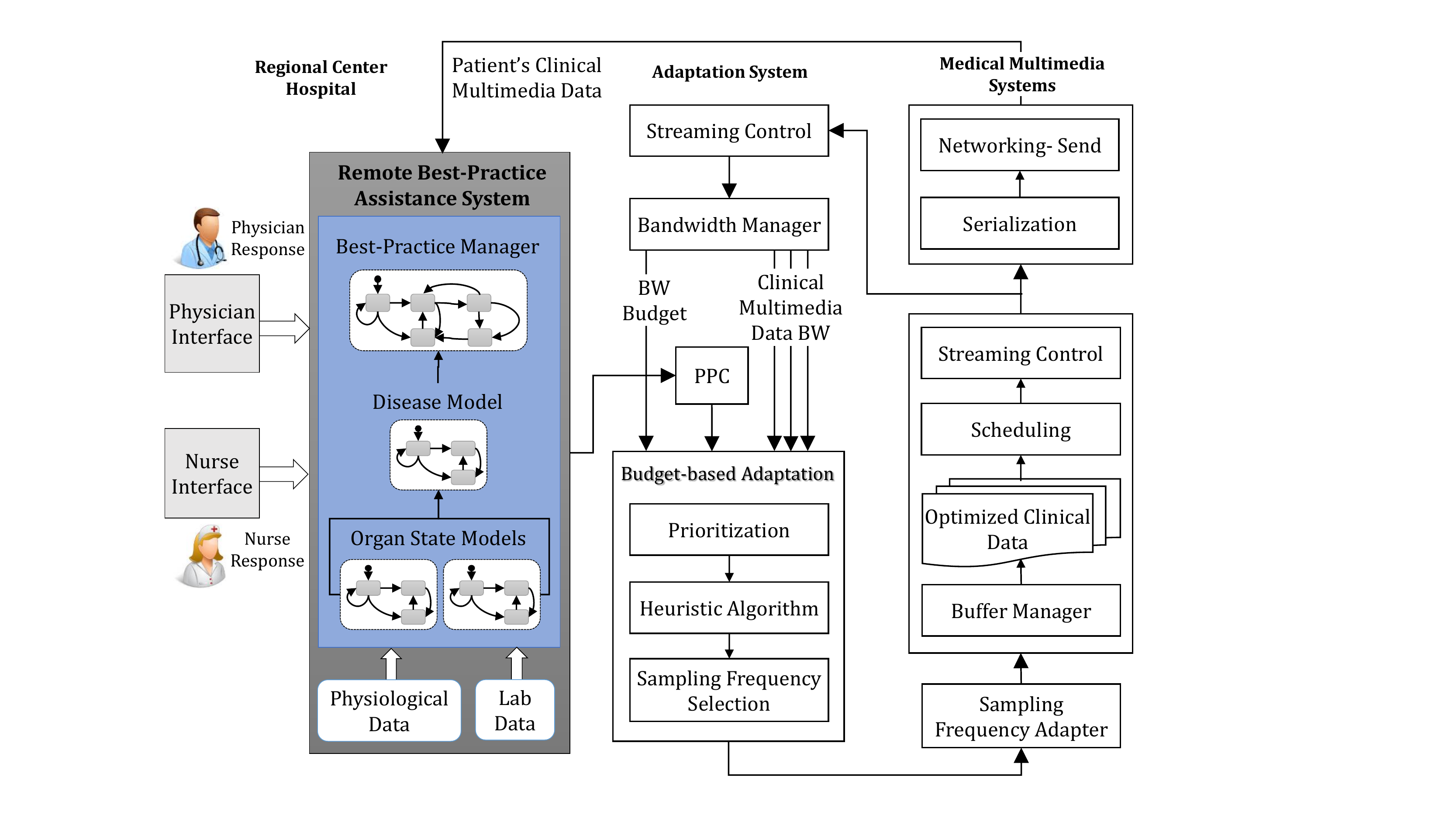}
\caption{Overview of our proposed system. Clinical data generated at the ambulance are adapted based on available bandwidth and inputs from a regional center hospital.}
\label{framework}
\vspace{-.3cm}
\end{figure*}

The design of our bandwidth-compliant physiology-aware adaptation system allows for efficient communication of clinical multimedia data in regards to the priority inputs from various clinical models while satisfying a bandwidth budget available to the ambulance. Figure \ref{scheme} shows a visual view of our model-driven prioritized adaptation system, showcasing an ambulance en route from a rural hospital to a regional center hospital communicating with the center hospital under a variable and limited bandwidth. The figure illustrates the notion of how different clinical states within an integrated set of various interacting clinical models can potentially change the priority of necessary clinical multimedia data within two consequent clinical states. It therefore adapts the bandwidth requirement of clinical multimedia to the available bandwidth. Figure \ref{manager} shows an abstract overview of our care management process feeding our adaptation system. All necessary clinical multimedia data including vital signs, audio-visual streams, information of neurological symptoms, and updates of disease states are monitored by doctors remotely in the center hospital, new patient physiological state is confirmed, and priority of each clinical multimedia data associated with current clinical state is then updated periodically.

The design of our bandwidth-compliant adaptation system revolves around the notion of model-driven physiology-aware adaptation and a two-level clinical multimedia data prioritization. The priority of each category of clinical multimedia data is considered in regards to the physiological models and the current state of underlying disease and organ models. A first-level priority is associated with each clinical multimedia data which depend on the disease models and its underlying organ models, and a second-level priority which depends on the state within each model. Under this prioritization schem, the adaptation system differentiates among clinical multimedia streams according to their criticality and select suitable adaptation per clinical multimedia stream for dynamic sampling frequency adjustment and bandwidth management. We adopt a Physiology-aware Priority Calculator (PPC) module which provides priority of clinical multimedia data at each point from inputs of physicians and nurses at a regional center hospital.

Figure \ref{framework} shows an overview of various modules and processes within our physiology-aware adaptation system. As a part of the adaptation system, the bandwidth manager module is devised as the core of the communication model to determine the current bandwidth budget and the bandwidth contribution of each specific clinical data stream. The information resulted from the bandwidth manager module is then fed to the main adaptation module. The first-level priority and second-level priority are integrated together and contribute to a global priority. The bandwidth budget, the clinical data streams bandwidth feed, along with the global priority are then used to provide adaptation by the heuristic algorithm that is described in the next subsection. Under the budget-based adaptations, proper sampling frequency of each clinical multimedia stream for all medical multimedia systems that generate clinical multimedia data (including medical devices and physiological systems) are selectively calculated and is sent to the sampling frequency adapter module for each medical system. The sampling frequency adapter then applies the specified sampling frequencies to the medical systems' buffer manager. The resulting patient clinical multimedia data are then compressed, encoded, scheduled for transmission, serialized, and streamed to the remote center hospital through the ambulance communication gateway to be finally decompressed, decoded, and remotely monitored by the physicians at the center hospital. 

The best-practice assistance system including the best-practice automata and physiological automata including disease and organ state models are run given the patient clinical multimedia data inputs received from the en-route ambulance. The PPC module is then derived by the recommendations from physicians remotely supervising the EMT at the ambulance. The cycle continues and in parallel the bandwidth manager monitors the variations in the bandwidth budget and the bandwidth contribution of each clinical multimedia stream. Transmission of an optimal and efficient bitrate not only reduces the communication bandwidth requirement, but also less exhausts other limited resources and reduces the scheduling and transmission delays from buffer, to process, and to model-driven visualization at the remote center hospital- a fact which is critical especially in remote supervision and emergency care.

\subsection{Problem Modeling}
For selection and transmission of clinical multimedia during emergency ambulance transport, one of the most significant factors in bandwidth usage is the sampling frequency of clinical multimedia which corresponds to the volume of multimedia data produced for the purpose of remote patient supervision. As discussed in Section \ref{sec:introduction}, communication bandwidth can get very limited in high-speed ambulances transferring a patient from a rural hospital to a regional, center hospital. Traveling through rural areas further exacerbates the problem due to the limited cellular data coverage. As discussed in Section \ref{sec:scenarios}, if the total bandwidth requirement of the all clinical multimedia data does not exceed the available bandwidth at any given time, then all of them can be transmitted as per standard protocol. If the available bandwidth budget \textit{W} is insufficient to transmit all the clinical multimedia streams, then an adaptation technique must be employed to reduce the bandwidth requirement of clinical multimedia data. Each category of clinical telemetry medical system consumes a specific bandwidth at any given time, and we must decide how to adopt a proper sampling frequency for each multimedia data type given their physiological state-dependent criticality so to be able to transmit the clinical multimedia streams within the available budget \textit{W}.

One approach for reducing the bandwidth requirement of communicating clinical multimedia data is to transmit a subset of the required clinical data. This selection scheme is the well-known binary Knapsack optimization problem. The binary Knapsack problem is NP-hard but efficient approximation algorithms can be utilized (fully polynomial approximation schemes), so this approach is computationally feasible. However, this method enables to only select a \textit{subset} of the necessary clinical multimedia data types, which is not safe in the emergency care scenario as \textit{all} the necessary multimedia streams and physiological information must be transmitted. Our proposed algorithms select all necessary clinical multimedia streams, but each with different bandwidth requirements according to their priorities. This is a multiple-choice knapsack problem in which the items (streams) are organized into groups corresponding to the objects. Each group contains the full-bandwidth clinical multimedia stream corresponding to an object and lower-bandwidth versions of the same clinical multimedia streams obtained by applying sampling frequency adaptation. Our proposed heuristics are all approximation algorithms for the multiple-choice knapsack problem, where the goal is to adapt to one proper clinical multimedia stream from each group within a bandwidth budget $W$ available to an en-route high-speed ambulance.

Different variations of the knapsack problem has been applied to certain contexts. Y. Song \textit{et al.} in their paper \cite{4753629} investigated the \textit{multiple} knapsack problem and its applications in cognitive radio networks. X. Xie and J. Liu in their paper \cite{4223174} studied \textit{quadratic} knapsack problem, while in \cite{6848188}, the authors applied the concept of \textit{m-dimensional} knapsack problem to the packet-level scheduling problem for a network, and proposed an approximation algorithm for that. Also, in our previous work \cite{storm}, we studied a \textit{quadratic 3-dimentional multiple} knapsack problem, and applied it against a complex NP-hard problem in the context of resource-aware stream processing.

\subsection{Adaptation of Clinical Data}
\label{sub:adapt}
Generally two classes of adaptation methods can be used to address the problem of bandwidth efficiency for the transmission of clinical multimedia data: active, and passive adaptations. Active adaptations focus on the scheduling algorithms of the 4G, 3G, or 2G mobile cellular networks to improve the bandwidth allocation for clinical multimedia data. However, the passive methods are inherently bandwidth-compliant adaptations with a focus to adapt the multimedia data transmission to the available bandwidth. We employ the second approach for the purpose of bandwidth-compliant clinical multimedia communication.

As discussed in Section \ref{sec:scenarios}, one of the important factors to consider clinical multimedia data communication is the criticality of any type of clinical multimedia data at any clinical state. The first step in our adaptation approach is to establish the criticality of each clinical multimedia data type within a specific physiological state. As mentioned earlier, the first-level priority is associated with each clinical multimedia data which depend on the disease models and its underlying organ models, and the second-level priority signifying the criticality of the state within each model. In this pilot study, we select a binary variable for the first-level priority signifying if the clinical state has higher or lower priority, and likewise a binary variable to account for the second-level priority signifying if the clinical multimedia data associated with that clinical state has higher or lower priority. Given this scheme, the PPC module classifies the required clinical data types into four clinical classes of pairs (first-level priority, second-level priority), namely $C_{11}$, $C_{12}$, $C_{21}$, and $C_{22}$, with $C_{11}$ representing the ``low", $C_{12}$ and $C_{21}$ the ``medium", and $C_{22}$ the ``high" clinical priority classes.

The optimization problem is to transmit the necessary set of clinical multimedia data in a way that maximizes a total of defined quality of clinical multimedia data as a quantitative measure of clinical effectiveness within an available bandwidth limit. The bandwidth requirement of a specific clinical multimedia data stream can be decreased by reducing the capturing sampling frequency. However, as per definition, this also results in a lower quality of that clinical multimedia data type. To take this into account, we ask physicians to define a maximum sampling frequency scaling called $R_{max}$ which is the maximum reduction in clinical multimedia data sampling frequency that is acceptable for the remote physicians. In this initial study, we assume that the quality of clinical multimedia data for a specific clinical multimedia stream is a function of its bandwidth requirement (with maximum quality of clinical multimedia data corresponding to minimum sampling frequency reduction) and its global priority. Without loss of generality and to enable a robust quantitative modeling in the simulation environment, we use the simplest function in this pilot study as the measure of quality of clinical multimedia: the product of consumed bandwidth and the global priority of clinical multimedia. For example, a clinical data stream $\tau_i$ of original bandwidth $s_{\tau_i}$ with sampling frequency scaling factor $r_i$ (therefore, the resulting sampling frequencies are factorials of the medical system's sampling rate and require no interpolation processing while sending data), and global priority of $p_{\tau_i}$ contributes to the quality of clinical multimedia data $s_{\tau_i}\times p_{\tau_i}\times r_i$. Our objective here is to apply sampling frequency scaling for sets of clinical multimedia data in a way that maximizes the total quality of clinical multimedia data subject to the constraints of the bandwidth budget $W$ at any given time and maximum sampling frequency reduction $R_{max}$.

\subsection{Proposed Solutions}
\label{sub:sol}
\begin{algorithm}[!t]
\begin{algorithmic}
\State $\mathcal{T}$: prioritized list of clinical multimedia data sorted from smallest to largest global priority
\State $\tau_i$: clinical multimedia data with original bandwidth requirement of $s_{\tau_i}$
\State $x_i$: adapted clinical multimedia data with bandwidth requirement of $s_{x_i}$
\State $R_{max} = 1/c^k$: maximum reduction of sampling frequency
\State Calculate $W_0 = \sum s_{\tau_i}\times R_{max}$ \%comment: minimum bandwidth requirement of all clinical multimedia data
\State $\forall \tau_i \in \mathcal{T}: s_{x_i} \gets s_{\tau_i}\times R_{max}$ \%comment: apply $R_{max}$ to all $\tau_i$'s.

\While {$s_{\tau_i}\times (1-R_{max}) \leq W_{i-1}$} \\
\%comment: i=1 initially.
\State $s_{x_i} \gets s_{\tau_i}$
\State $W_i \gets W_{i-1} - s_{\tau_i}\times (1-R_{max})$
\State{ $i \gets i+1$~~\%comment: adapt the sampling frequency of next clinical multimedia data}
\EndWhile \\
\%comment above loop repeats until some clinical multimedia data $\tau_{\ell}$ cannot be received at original bandwidth within the remaining bandwidth budget $W_{\ell-1}$.
\State $\ell \gets i$ \%comment: resulting from above loop.
\State{ Find minimum $r_{l}=1/c^l$ such that \\$s_{\tau_{\ell}}\times r_{l}\leq W_{\ell-1}+ (s_{\tau_{\ell}}\times R_{\max})$ ~~\%comment: determine the maximum bandwidth at which $\tau_{\ell}$ can e received by calculating the minimum bandwidth reduction $r_{\ell}$.}
\State{ $s_{x_{\ell}} \gets s_{\tau_{\ell}}\times r_{l}$ \%comment: adapt $\tau_{\ell}$ and calculate $s_{x_{\ell}}$}
\end{algorithmic}
 \caption*{Algorithm 1: Compromise}
\end{algorithm}

There are $n$ clinical data types $\mathcal{T}=\{\tau_1,\tau_2,\ldots,\tau_n\}$, and each $\tau_h \in \mathcal{T}$ has an original bandwidth requirement of $s_{\tau_h}$, and a global priority or criticality $p_{\tau_h}$. The quality of clinical multimedia data $\tau_h$ is $q_{\tau_h} = p_{\tau_h}\times s_{\tau_h}$. The total travel time for which adaptations are done is $T$, and the available bandwidth limits the total bandwidth requirement of the clinical multimedia data that must be transmitted to the regional center hospital to $W$.

Let $X= \{x_1,x_2,\ldots,x_n\}$, be the set of clinical data that are transmitted to the regional center hospital. Each $x_i \in X$ corresponds to an original clinical multimedia data $\tau_i \in \mathcal{T}$ with a global priority $p_{x_i} = p_{\tau_i}$. The bandwidth requirement of a specific clinical multimedia stream $x_i \in X$ is the original bandwidth requirement of the clinical multimedia stream $\tau_i \in \mathcal{T}$ scaled by a factor of $c^{r_i}$ for some $0 \leq r_i \leq k$ and constant $c$ where $R_{max}=1/c^k$ is specified by the physicians as the \textit{maximum reduction of sampling frequency} clinically. Our choice of a scaling factor of $c^{r_i}$ was motivated by the systems based on Distributed Hash Table such as Chord \cite{chord} in which the distance between a node and its fingers increases exponentially. So, the bandwidth requirement of clinical data $x_i$ is $s_{x_i}=\frac{s_{\tau_i}}{c^{r_i}}$, and the quality of clinical multimedia data $x_i$ is $q_{x_i}=p_{x_i}\times s_{x_i}= \frac{p_{\tau_i}\times s_{\tau_i}}{c^{r_i}}$ as per our initial definitions. It should be noted that the clinical automata are codified and already available by the physicians. $\mathcal{T}$ and the priority $p_{\tau_i}$ of each individual clinical multimedia data $\tau_h \in \mathcal{T}$ within the context of that specific clinical state are given in advance. Whenever a change of clinical state is triggered, $\mathcal{T}$ is updated accordingly, and adaptation heuristic algorithm is run.

\subsection{Heuristic Algorithms}
\begin{algorithm}[!t]
\begin{algorithmic}
\State $\mathcal{T}$: prioritized list of clinical multimedia data sorted from smallest to largest global priority
\State $\tau_i$: clinical multimedia data stream with original bandwidth $s_{\tau_i}$
\State $x_i$: adapted clinical multimedia data stream with bandwidth $s_{x_i}$
\State $R_{max} = 1/c^k$: maximum reduction of bandwidth
\While {$\sum s_{x_i} \leq W$}
\State Find maximum $j<k$ such that $r_{i}=1/c^j$ and \\$s_{\tau_{i}}\times r_{i}\leq W_{i-1}+ (s_{\tau_{i}}\times R_{\max})$ ~~\%comment: determine minimum bandwidth reduction $r_{i} \geq R_{\max}$
\State $s_{x_{i}} \gets s_{\tau_{i}}\times r_{i}$ ~~\%comment: adapt $\tau_{i}$ and calculate $s_{x_{i}}$
\State $i \gets (i+1)~\%~n$~~\%comment: adapt the sampling frequency of next clinical multimedia data; if the end, start from the beginning
\EndWhile
\end{algorithmic}
 \caption*{Algorithm 2: Round-Robin}
\end{algorithm}

Let $S$ be the total bandwidth requirement of transmitting all clinical multimedia streams, and $W$ be the available bandwidth at a given point. The physician-defined maximum reduction of sampling frequency is $R_{max} = 1/c^k$. Let $C_{11}$ be the class of clinical multimedia streams with the smallest global priority, and similarly for $C_{12}$, $C_{21}$, and $C_{22}$. For each clinical multimedia stream $\tau_i$ in the list, we calculate the quality of clinical data $q_i$ as described previously. This is the contribution that $\tau_i$ would make to the average quality of the clinical multimedia data for the whole adaptation period if it was received at original bandwidth. We then calculate $W_{\min} = S\times R_{\max}$ which is the minimum bandwidth requirement needed for transmission of all clinical multimedia streams. In the following, assume that $W_{\min} \leq W$ so the unused bandwidth budget is $W_0=W-W_{\min}$.

\begin{algorithm}[!t]
\begin{algorithmic}
\State $\mathcal{T}$: prioritized list of clinical multimedia data sorted from smallest to largest global priority
\State $\tau_i$: clinical multimedia data stream with original bandwidth $s_{\tau_i}$
\State $x_i$: adapted clinical multimedia data stream with bandwidth $s_{x_i}$
\State $R_{max} = 1/c^k$: maximum reduction of bandwidth
\While {$\sum s_{x_i} \leq W$}
\Repeat \State{ Find maximum $j<k$ such that $r_{i}=1/c^j$ and \\$s_{\tau_{i}}\times r_{i}\geq W_{i-1}+ (s_{\tau_{i}}\times R_{\max})$ ~~\%comment: determine minimum bandwidth reduction $r_{i} \geq R_{\max}$}
\State{ $s_{x_{i}} \gets s_{\tau_{i}}\times r_{i}$ \%comment: adapt $\tau_{i}$ and calculate $s_{x_{i}}$}
\Until{$j\leq k$} ~~\%comment: $r_{i}=R_{max}$
\State{ $i \gets i+1$~~\%comment: adapt the sampling frequency of next clinical multimedia data}
\EndWhile
\end{algorithmic}
 \caption*{Algorithm 3: Aggressive}
\end{algorithm}

To determine the bandwidth reduction for each clinical multimedia data, our main heuristic algorithm (namely \textit{Compromise} represented in Algorithm 1) sorts the prioritized list of clinical multimedia streams by their global priority from the largest to the smallest. For ease of notation in the following, suppose that the clinical data streams are re-indexed so that the sorted list of clinical multimedia data is $\tau_1,\tau_2,\ldots,\tau_n$. If $s_{\tau_1}\times(1- R_{\max})\leq W_0$ then there is enough unused bandwidth to receive $\tau_1$ at original full bandwidth, so the clinical data $x_1$ has $s_{x_1}=s_{\tau_1}$ and contributes $q_1$ to the average quality of clinical multimedia data. This leaves an unused bandwidth of $W_1=W_0-s_{\tau_1} \times (1- R_{\max})$ for the remaining clinical multimedia streams after $x_1$. The algorithm repeats for $\tau_2, \tau_3,\ldots$ until some clinical data $\tau_{\ell}$ cannot be transmitted at original bandwidth within the remaining budget $W_{\ell-1}$. It then determines the maximum bandwidth at which it can be transmitted by calculating the minimum bandwidth reduction $r_{\ell}$ which similar to $R_{max}$ is a multiplicative inverse (reciprocal) of a power of $c$ such that $s_{\tau_{\ell}}\times r_{\ell}\leq W_{\ell-1}+ (s_{\tau_{\ell}}\times R_{\max})$. The received clinical data $x_{\ell}$ will have bandwidth $s_{x_{\ell}} = s_{\tau_{\ell}}\times r_{\ell}$ and will contribute $q'_{\ell}$ to the average quality of clinical multimedia data of the whole adaptation period. The remaining bandwidth budget after transmission of $x_{\ell}$ will be $W_{\ell} = W_{\ell-1} - (s_{\tau_{\ell}}\times r_{\ell}))$. The algorithm repeats this process to determine the available bandwidth budget and quality contribution for each of the remaining clinical data $x_{\ell+1},x_{\ell+2},\ldots,x_{n}$. Finally the total quality of clinical multimedia data and other statistics are calculated.

The other two heuristic algorithms (namely \textit{Aggressive} and \textit{Round-Robin}) are also represented in Algorithm 2 and Algorithm 3 respectively. All our heuristic algorithms run in real-time and incur trivial overhead during the runtime due to the fact that the total number of clinical multimedia data and the number of priority classes are small- a fact necessary for real-time remote monitoring of patients. Our heuristics are implemented efficiently in $O(n log n)$ time and $O(n)$ space and produce solutions very close optimal. The approximation error depends on the difference between the bandwidth chosen for the first clinical data that cannot be received at original bandwidth and the remaining bandwidth available to transmit it. 
Also, it should be noted that the adaptive clinical scheme only changes the sampling frequency but not other metrics leading to bandwidth changes of clinical multimedia data (such as multimedia compression ratios). However, our proposed optimization scheme could be adopted in a similar way to controlling the compression ratio as well. 
\section{Evaluation}
\label{sec:evaluation}

\begin{figure}[!t]
\centering
\includegraphics[width=1.02\columnwidth, trim = 50 250 50 265, clip = true]{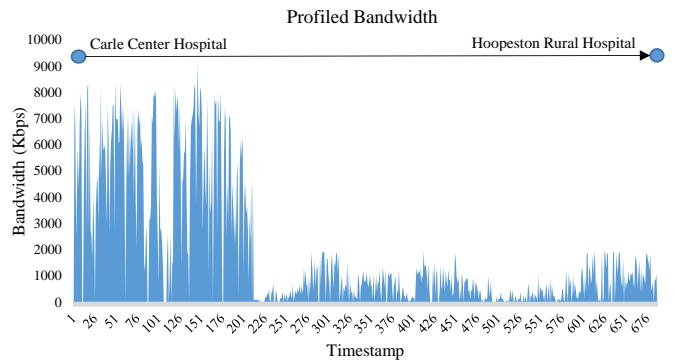}
\caption{A sample of profiled bandwidth rates along a route from Hoopeston rural hospital to Carle center hospital.}
\label{profile}
\end{figure}

\begin{figure*}[!p!t]
\centering
\includegraphics[totalheight=0.21\textheight, trim = 45 240 45 240, clip = true]{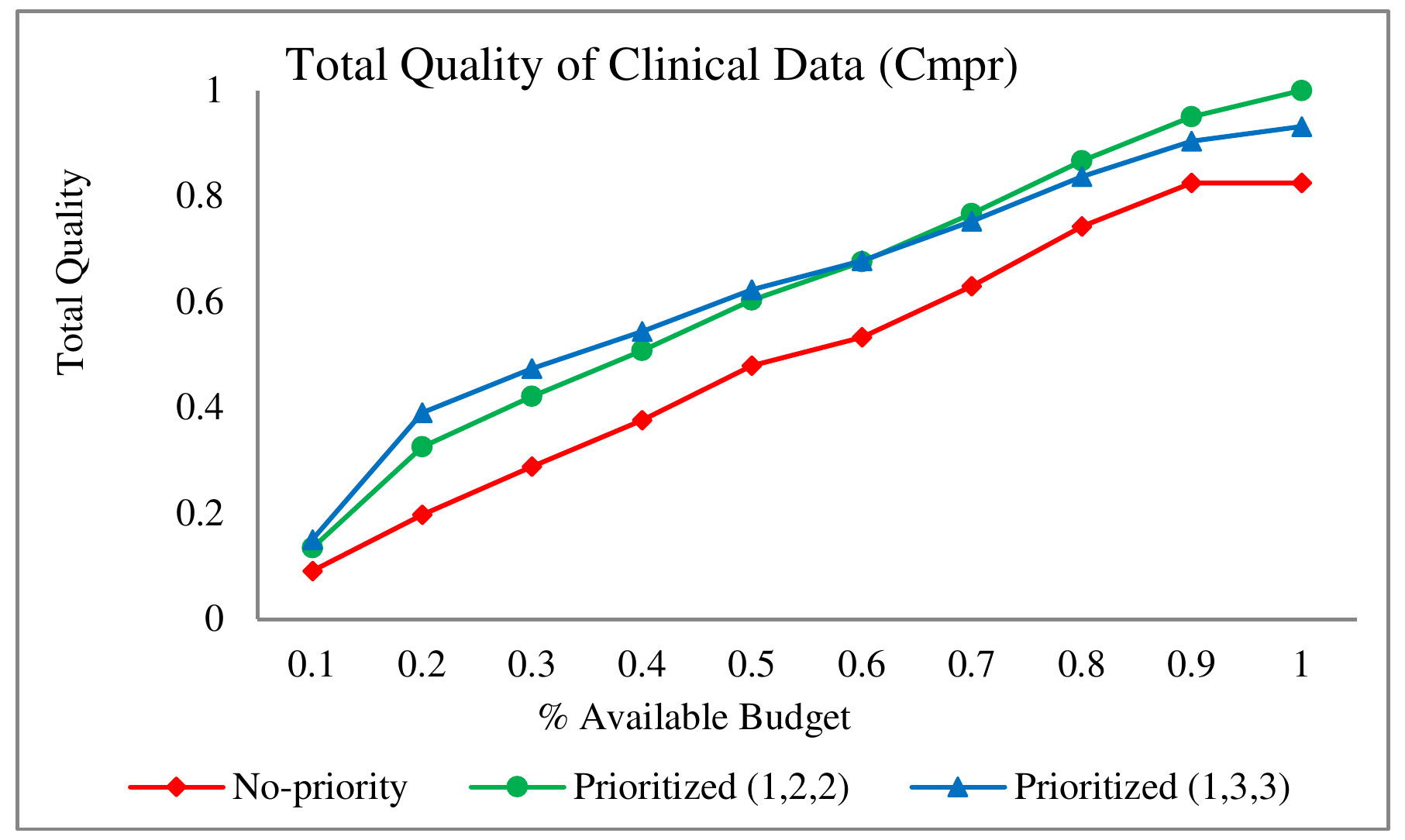}\includegraphics[totalheight=0.21\textheight, trim = 45 240 45 240, clip = true]{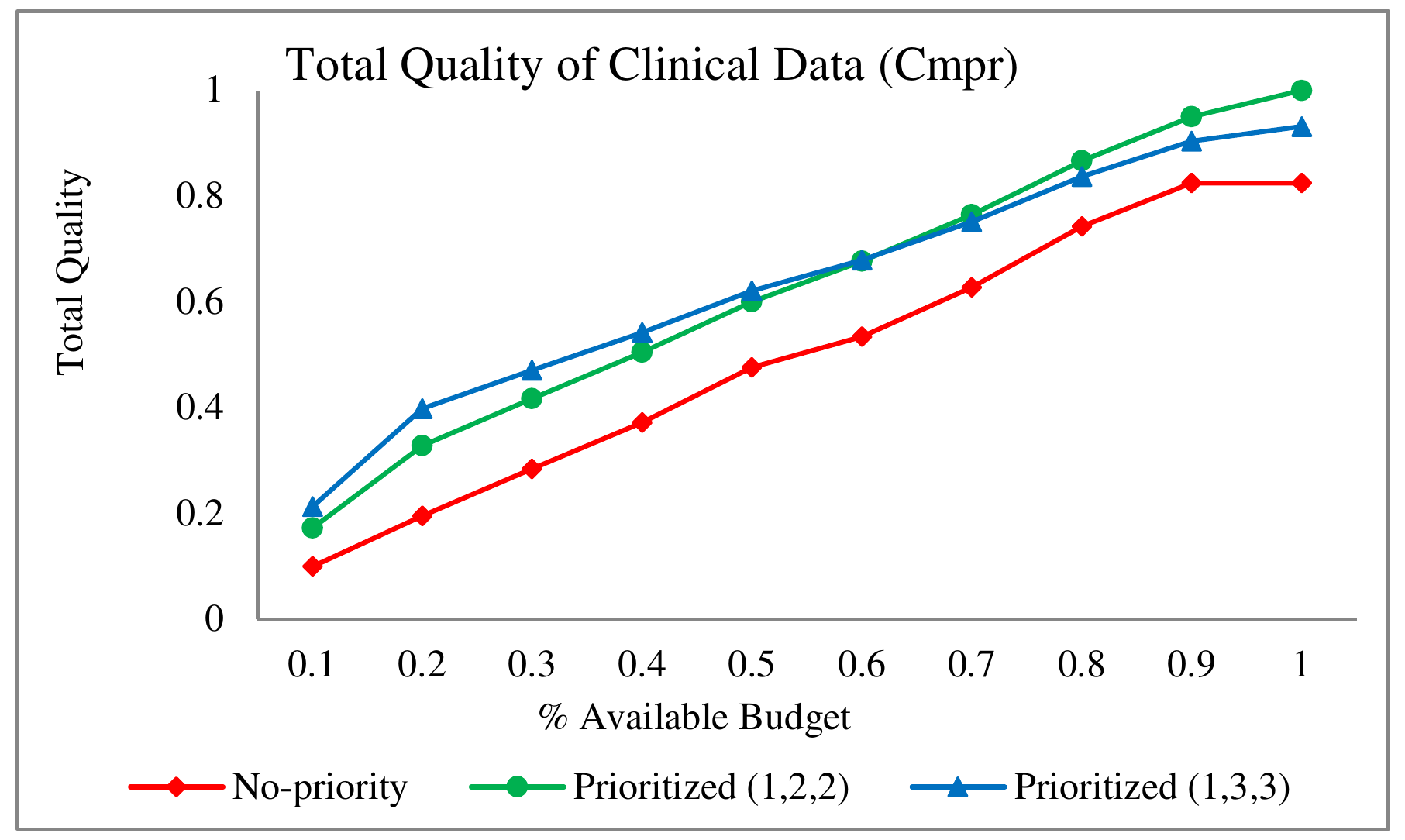}
\includegraphics[totalheight=0.21\textheight, trim = 45 240 45 240, clip = true]{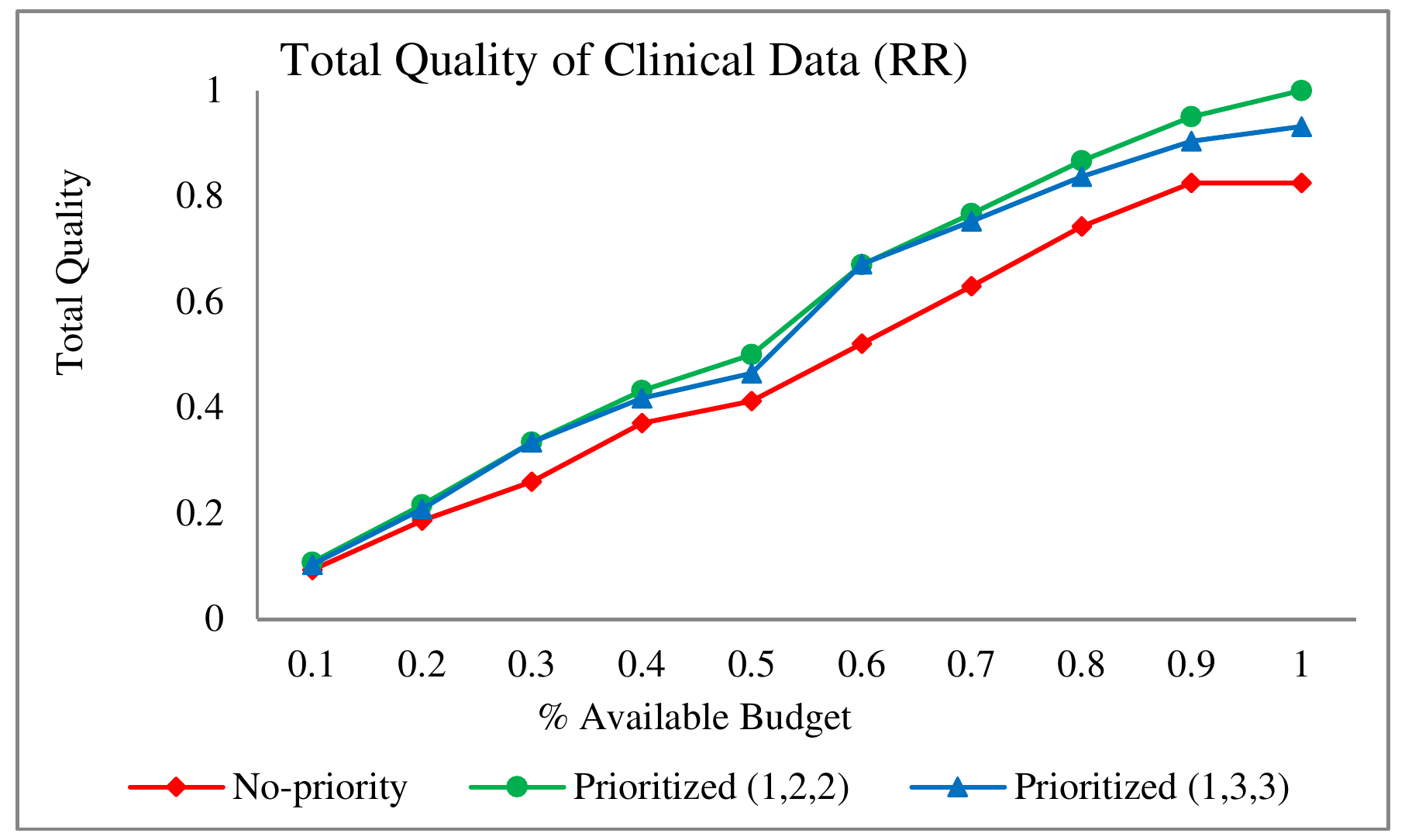}\includegraphics[totalheight=0.21\textheight, trim = 45 240 45 240, clip = true]{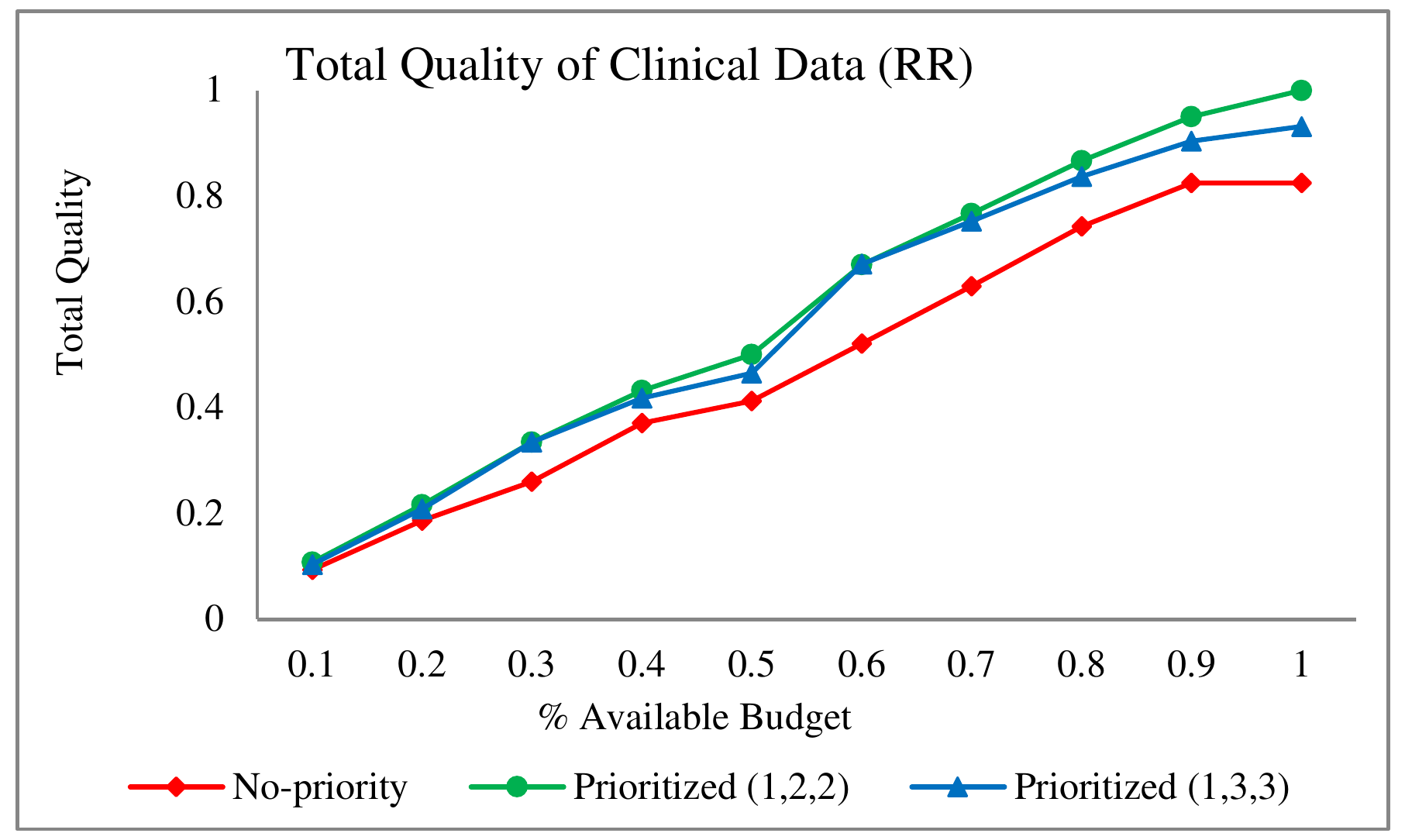}
\includegraphics[totalheight=0.21\textheight, trim = 45 240 45 240, clip = true]{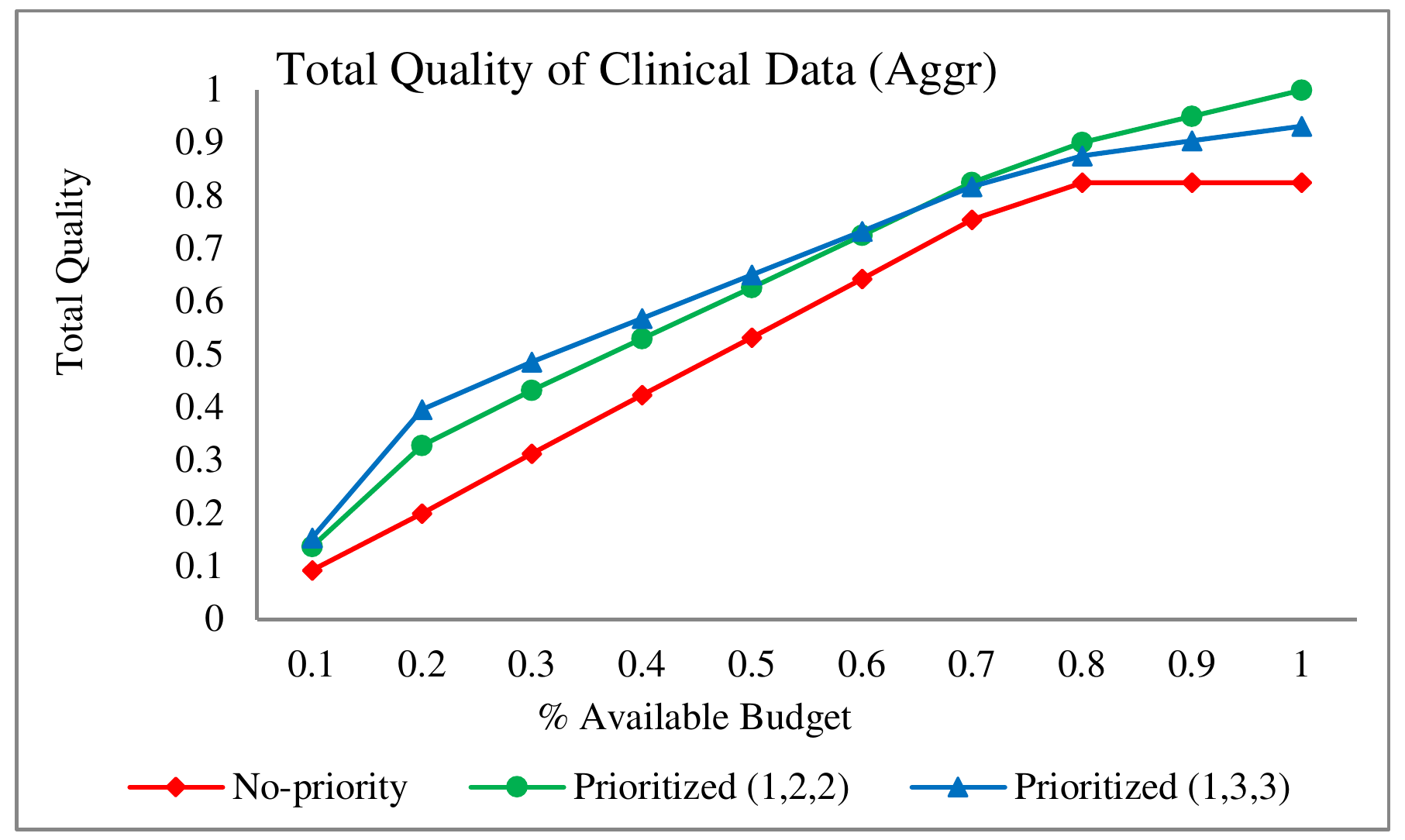}\includegraphics[totalheight=0.21\textheight, trim = 45 240 45 240, clip = true]{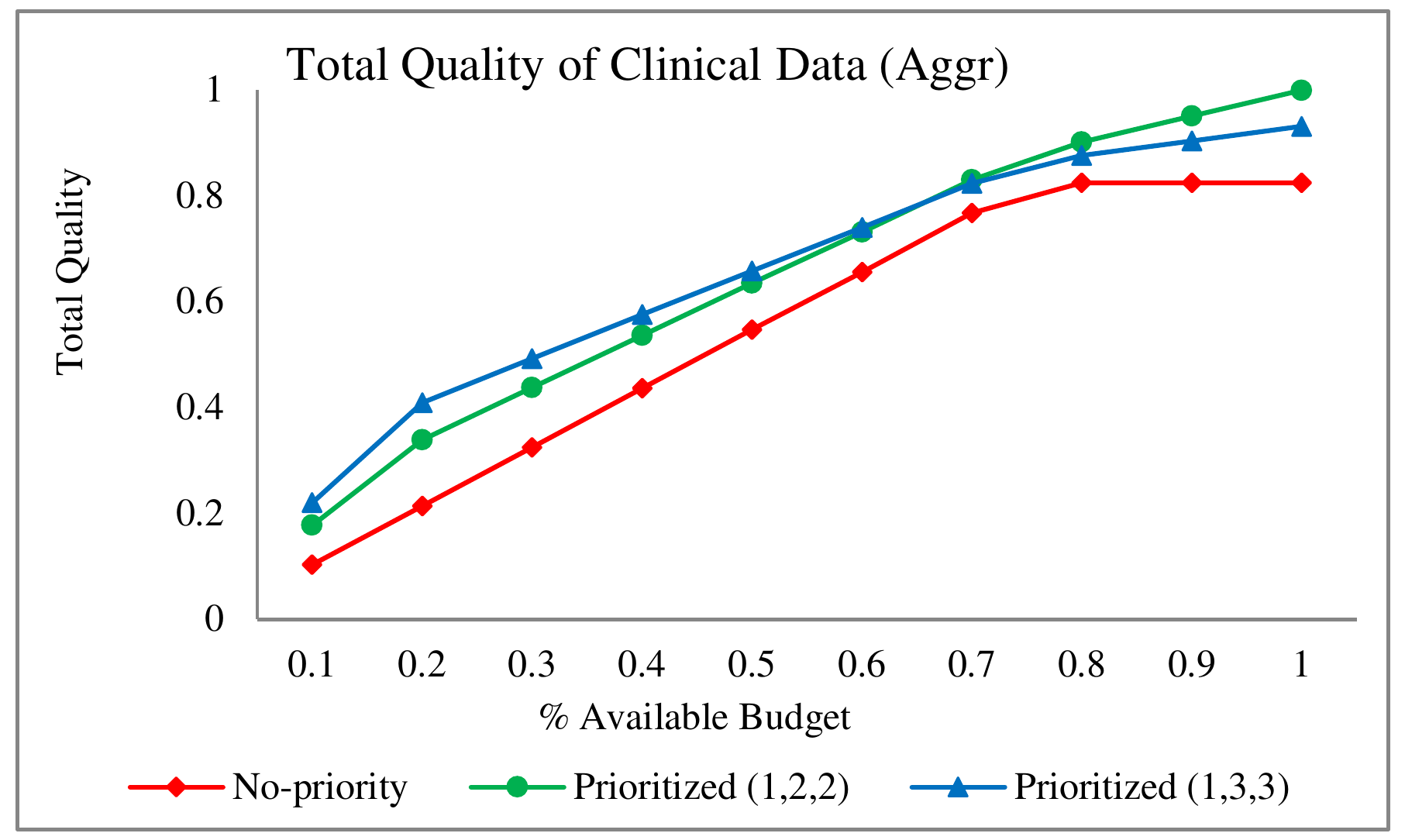}
\caption{Total quality of clinical multimedia data measured for \textit{Aggressive}, \textit{Round-Robin}, and \textit{Compromise} algorithms, for four priority classes, and two values of $R_{max}=1/c^k$. (Top-Left) Compromise, $k=4$, (Top-Right) Compromise, $k=6$, (Middle-Left) Round-Robin, $k=4$, (Middle-Right) Round-Robin, $k=6$, (Bottom-Left) Aggressive, $k=4$, (Bottom-Right) Aggressive, $k=6$.}
\label{pdf:quality}
\end{figure*}

\begin{figure*}[!p!t]
\centering
\includegraphics[totalheight=0.21\textheight, trim = 45 240 45 240, clip = true]{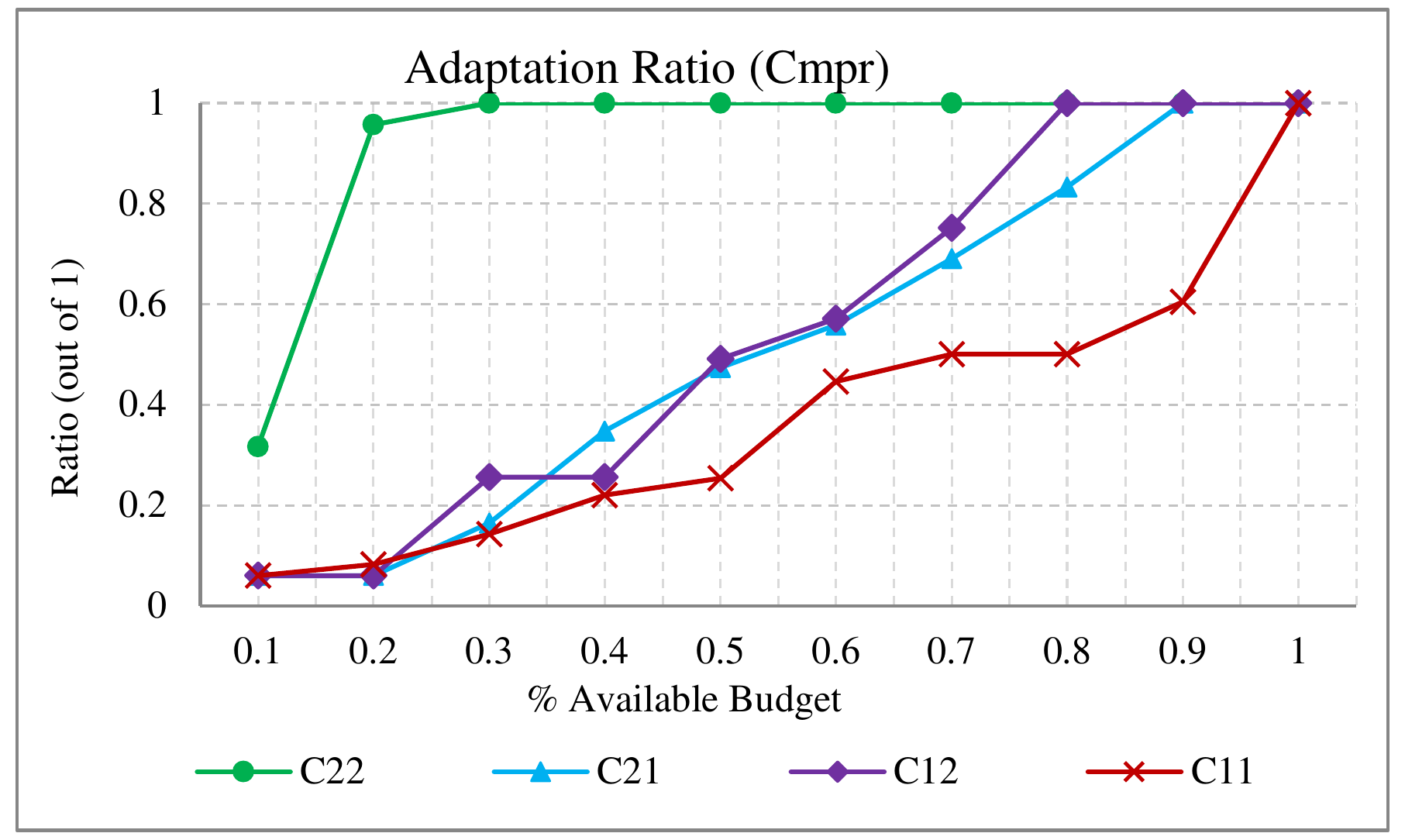}\includegraphics[totalheight=0.21\textheight, trim = 45 240 45 240, clip = true]{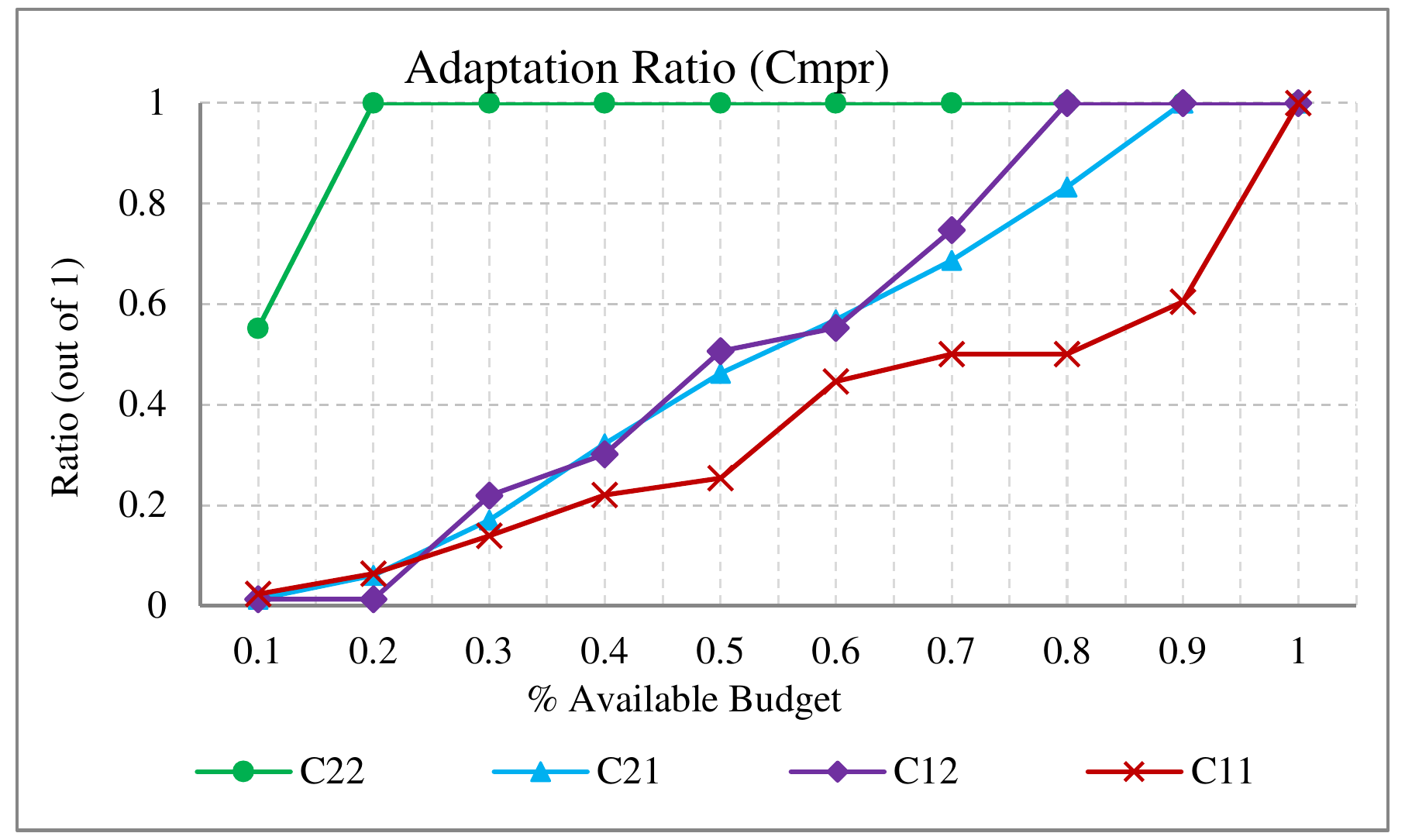}
\includegraphics[totalheight=0.21\textheight, trim = 45 240 45 240, clip = true]{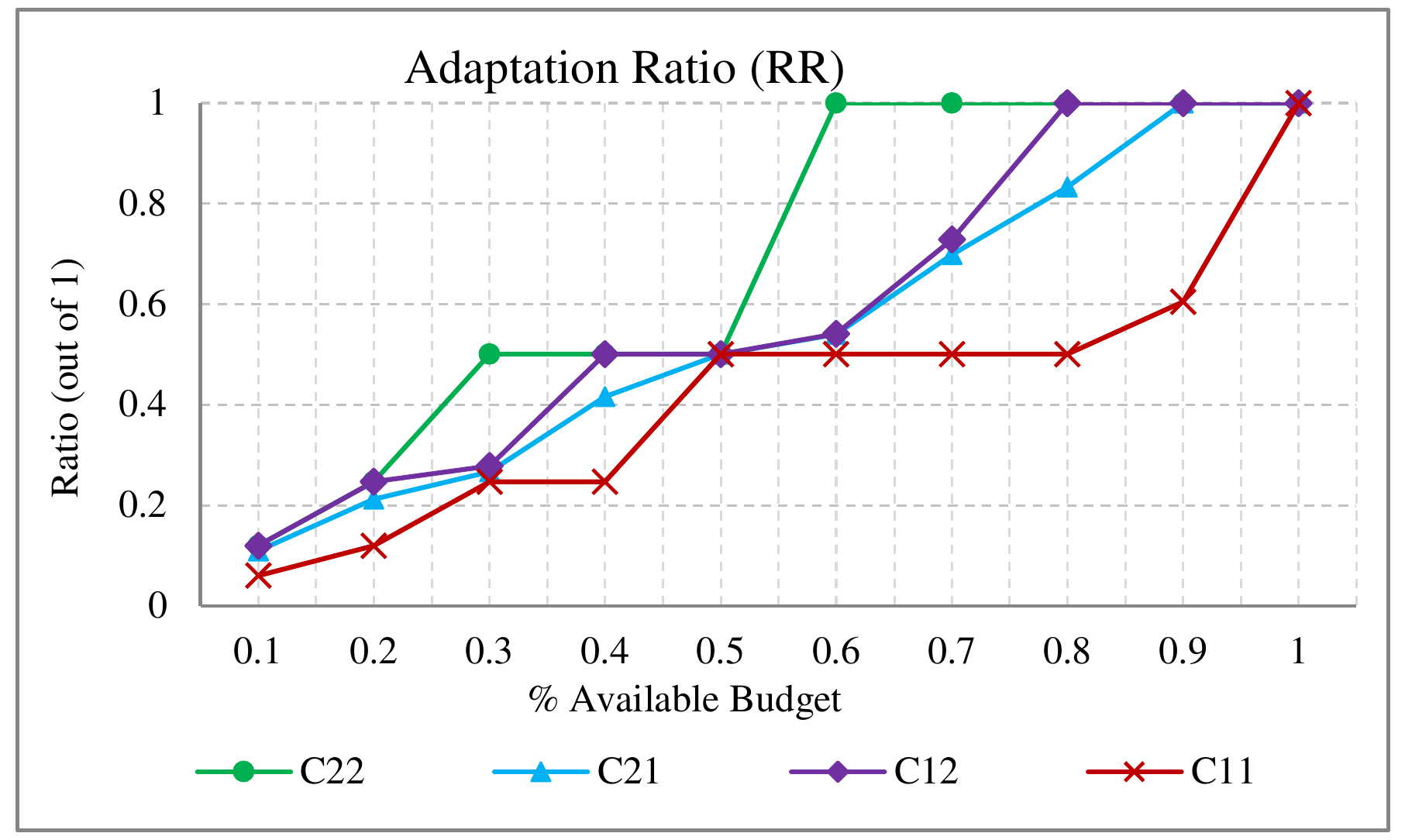}\includegraphics[totalheight=0.21\textheight, trim = 45 240 45 240, clip = true]{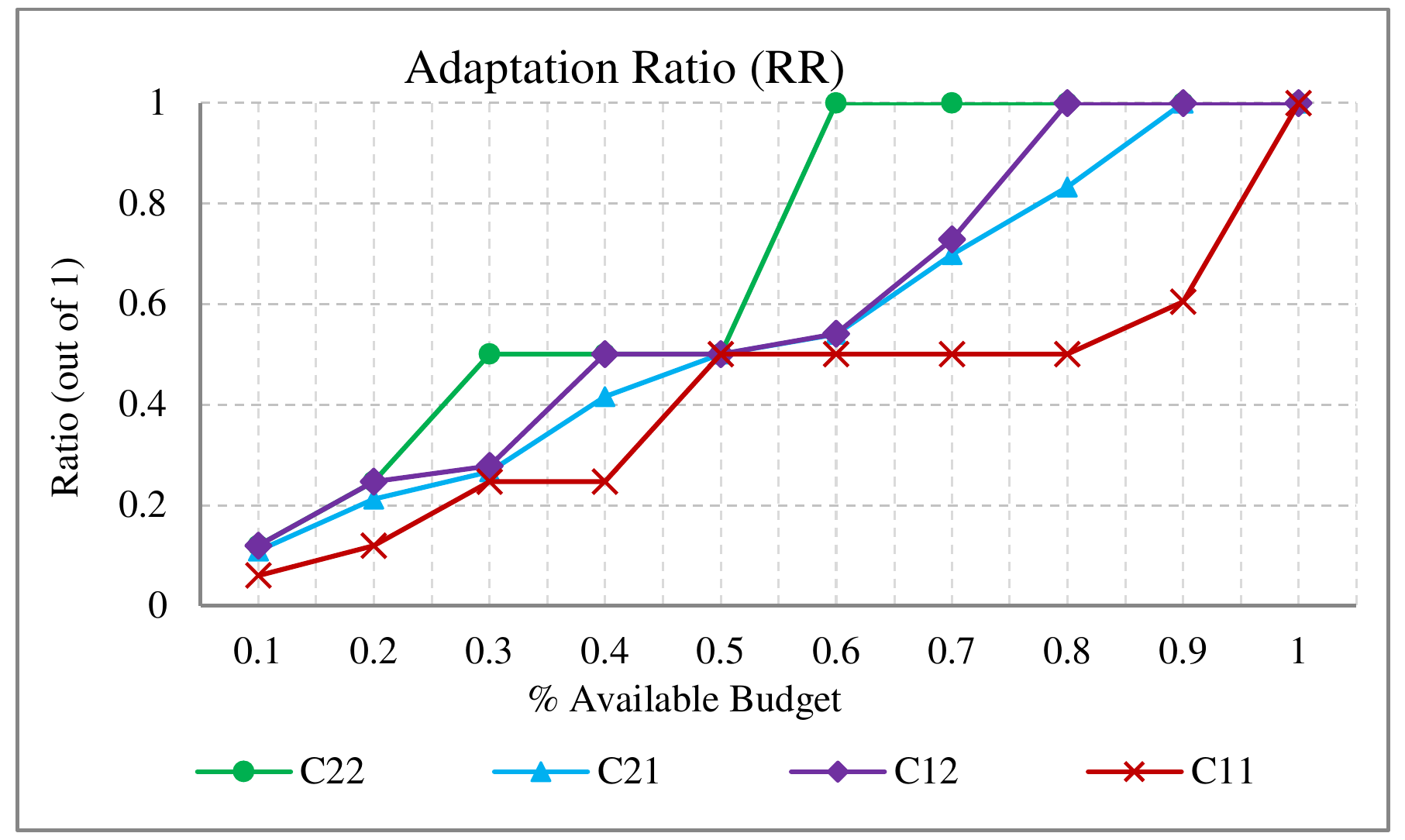}
\includegraphics[totalheight=0.21\textheight, trim = 45 240 45 240, clip = true]{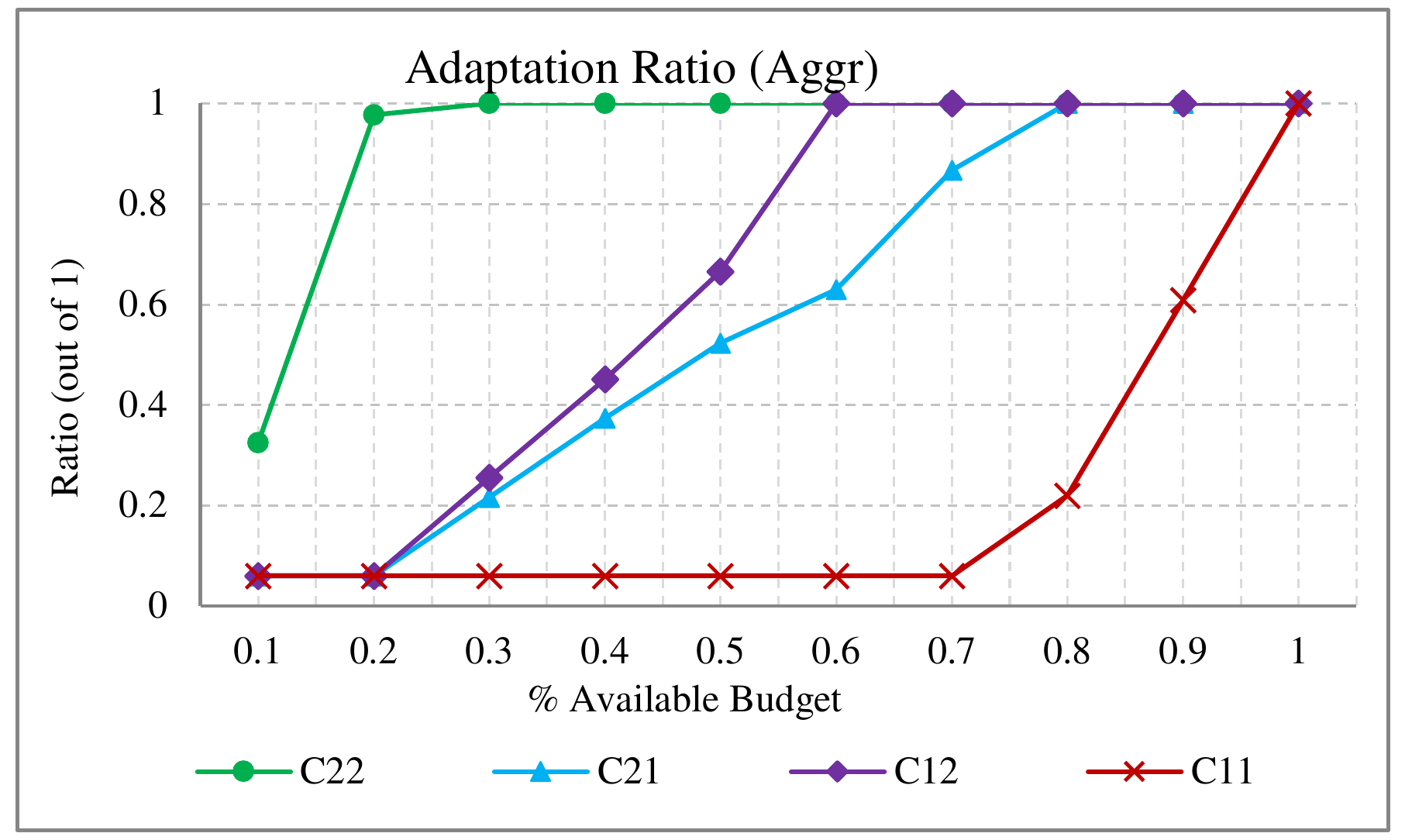}\includegraphics[totalheight=0.21\textheight, trim = 45 240 45 240, clip = true]{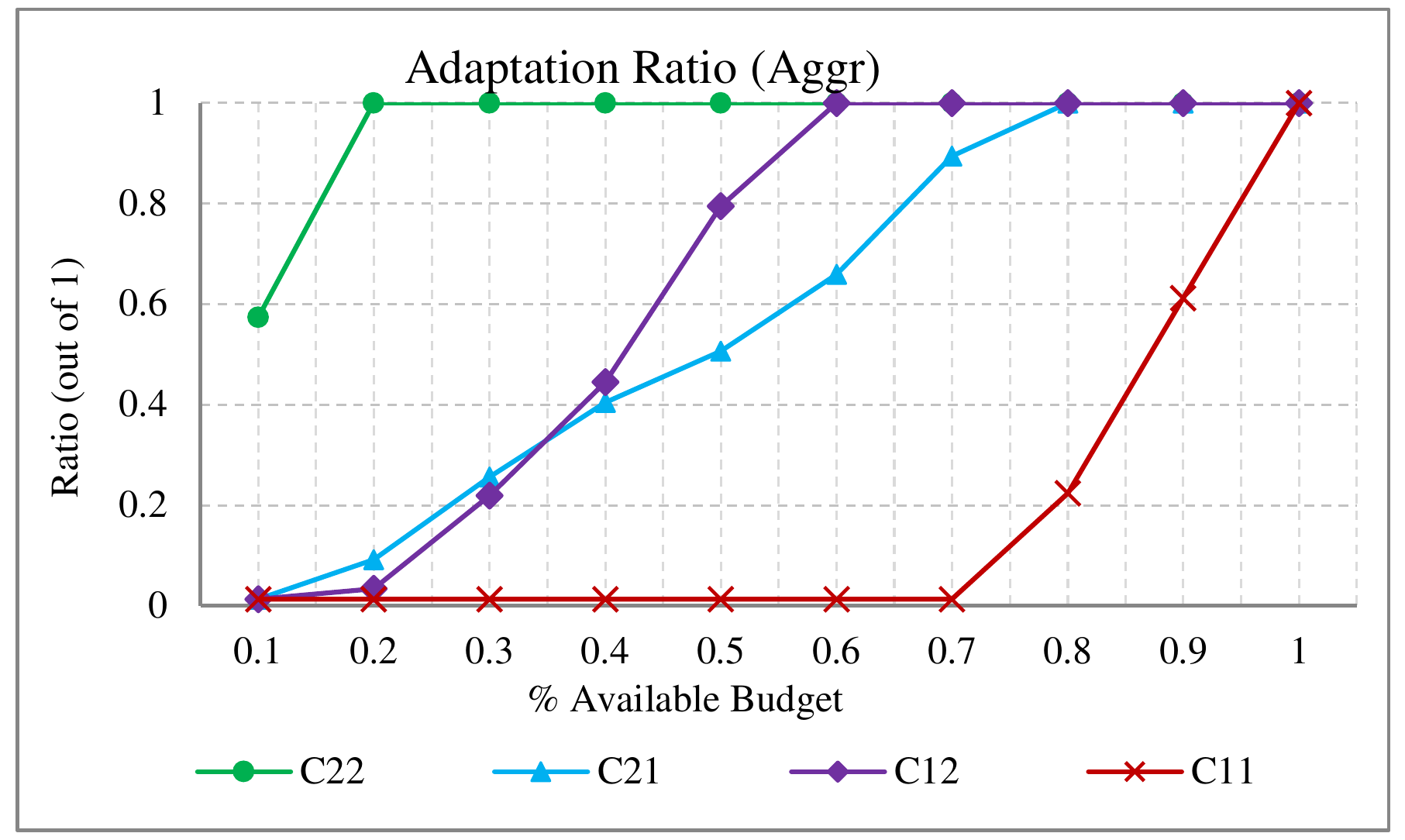}
\caption{Adaptation ratio in terms of bandwidth reduction, measured for four priority classes, and two values of $R_{max}=1/c^k$. (Top-Left) Compromise, $k=4$, (Top-Right) Compromise, $k=6$, (Middle-Left) Round-Robin, $k=4$, (Middle-Right) Round-Robin, $k=6$, (Bottom-Left) Aggressive, $k=4$, (Bottom-Right) Aggressive, $k=6$.}
\label{pdf:adaptations}
\end{figure*}

\begin{figure*}[!p!t]
\centering
\includegraphics[totalheight=0.21\textheight, trim = 45 240 45 240, clip = true]{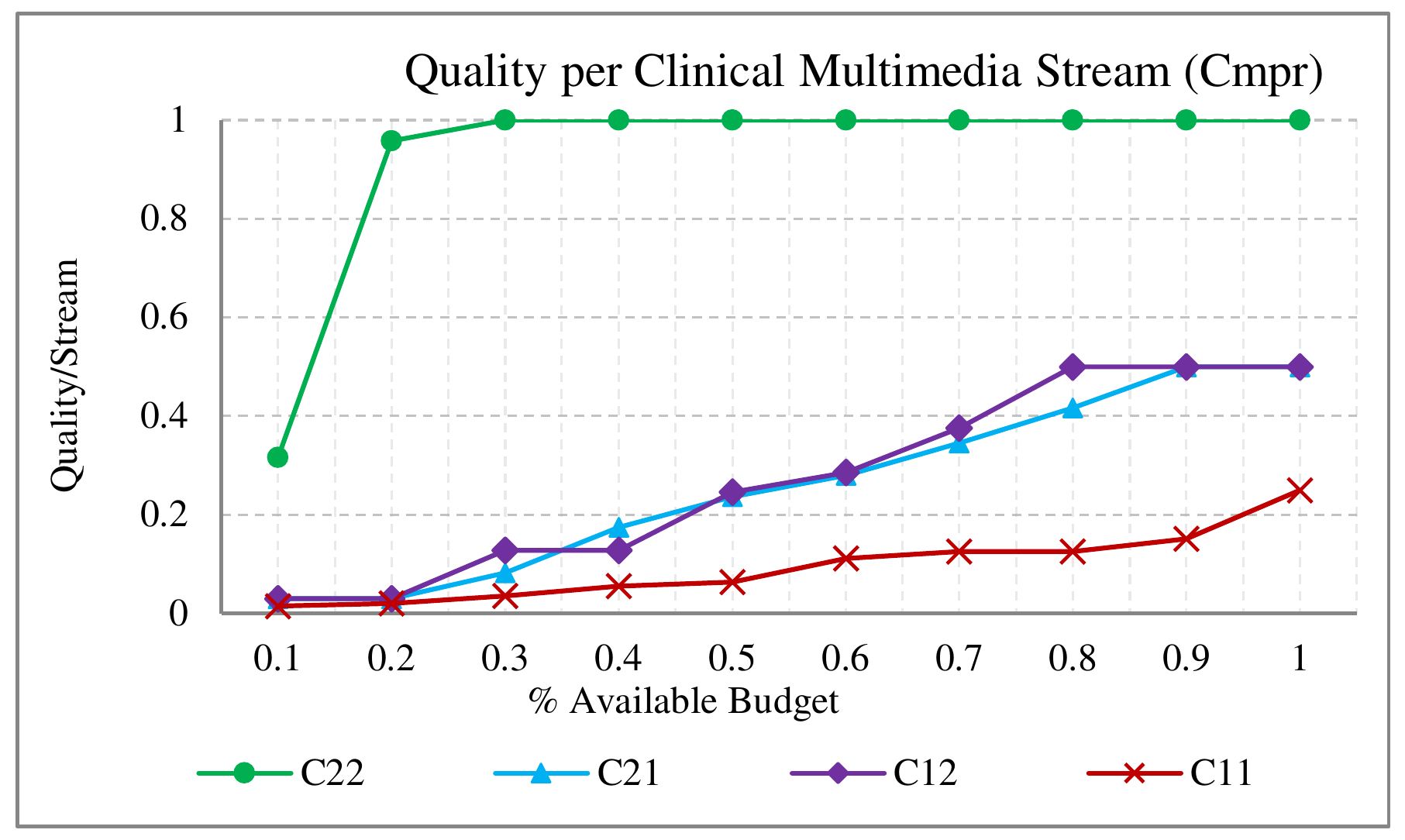}\includegraphics[totalheight=0.21\textheight, trim = 45 240 45 240, clip = true]{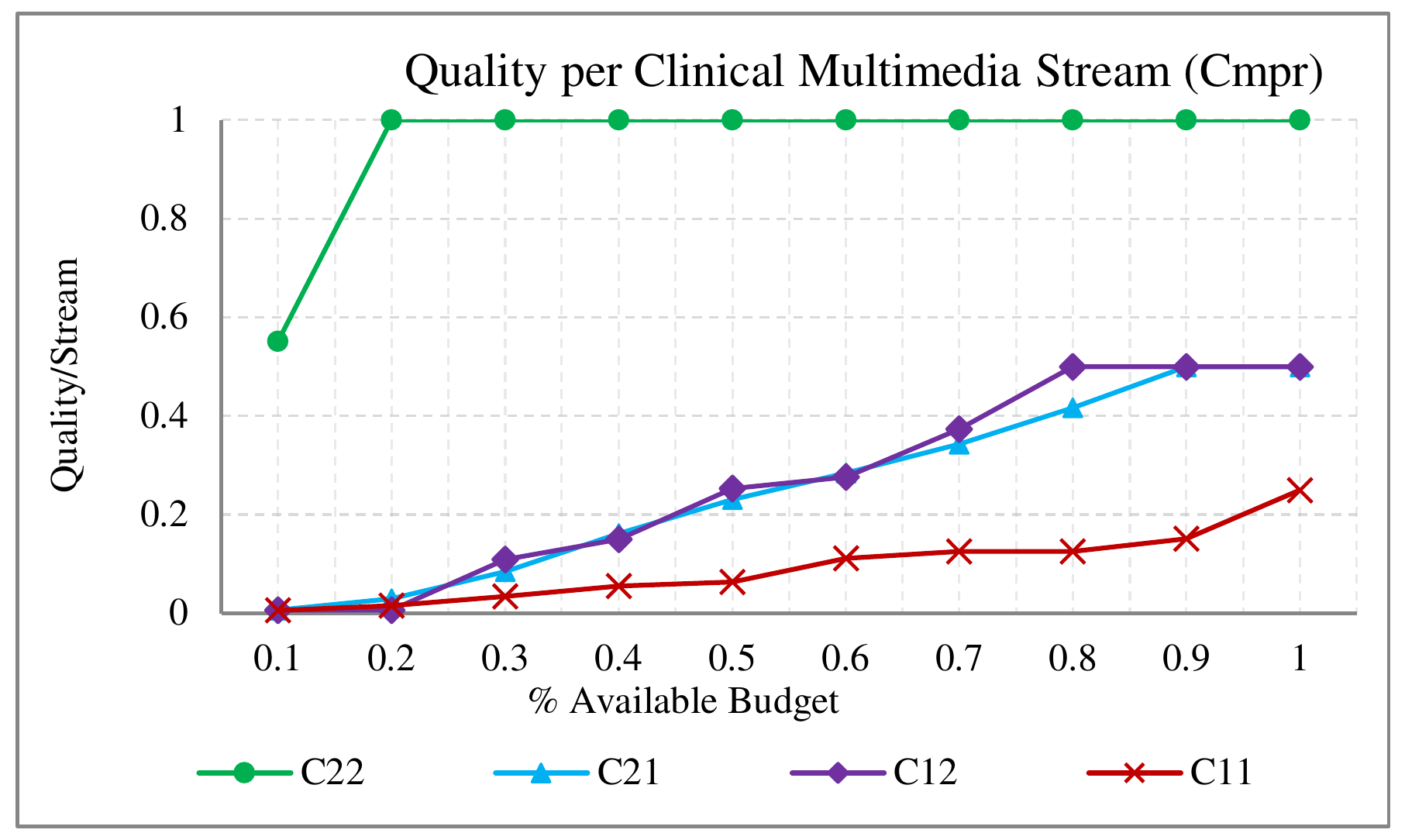}
\includegraphics[totalheight=0.21\textheight, trim = 45 240 45 240, clip = true]{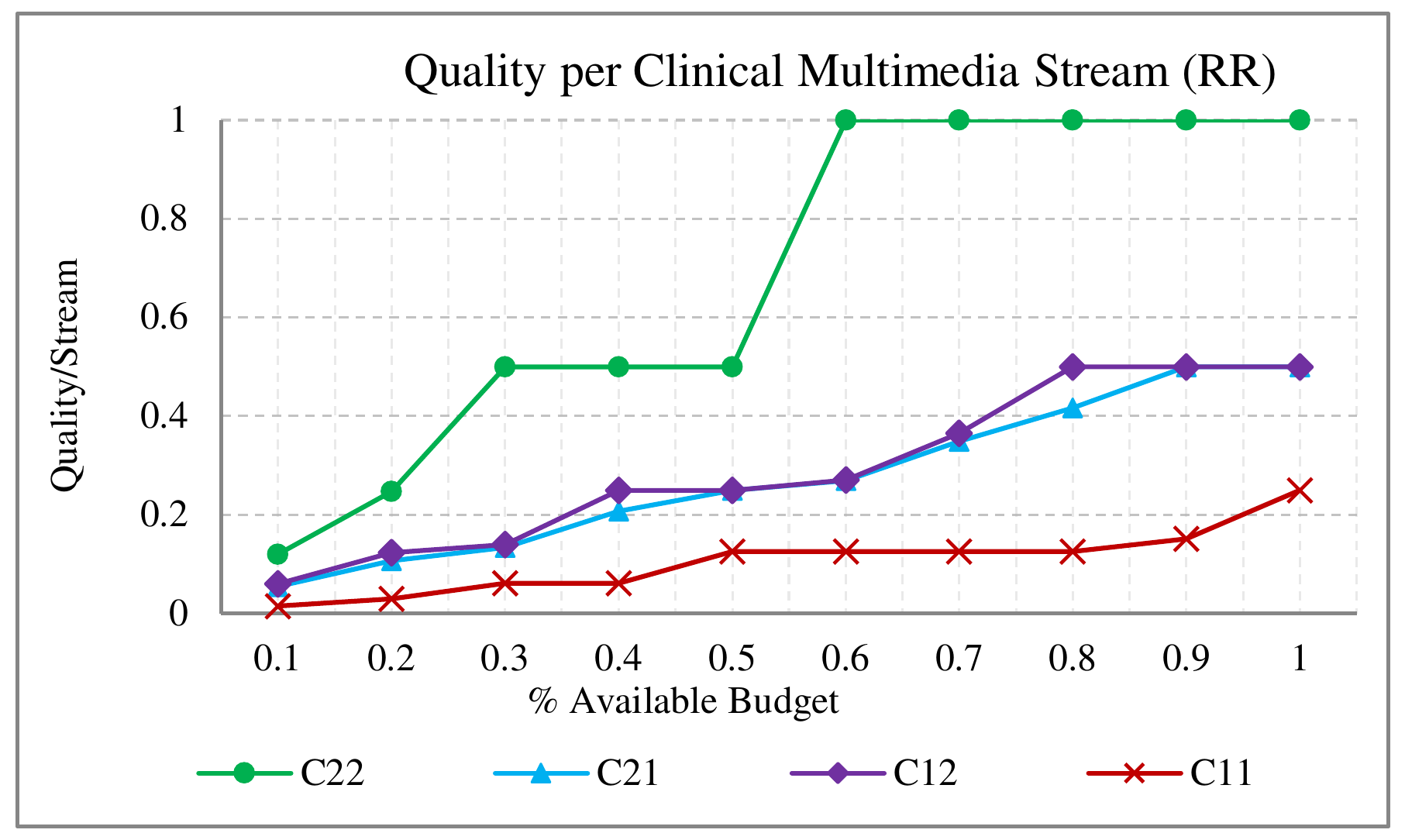}\includegraphics[totalheight=0.21\textheight, trim = 45 240 45 240, clip = true]{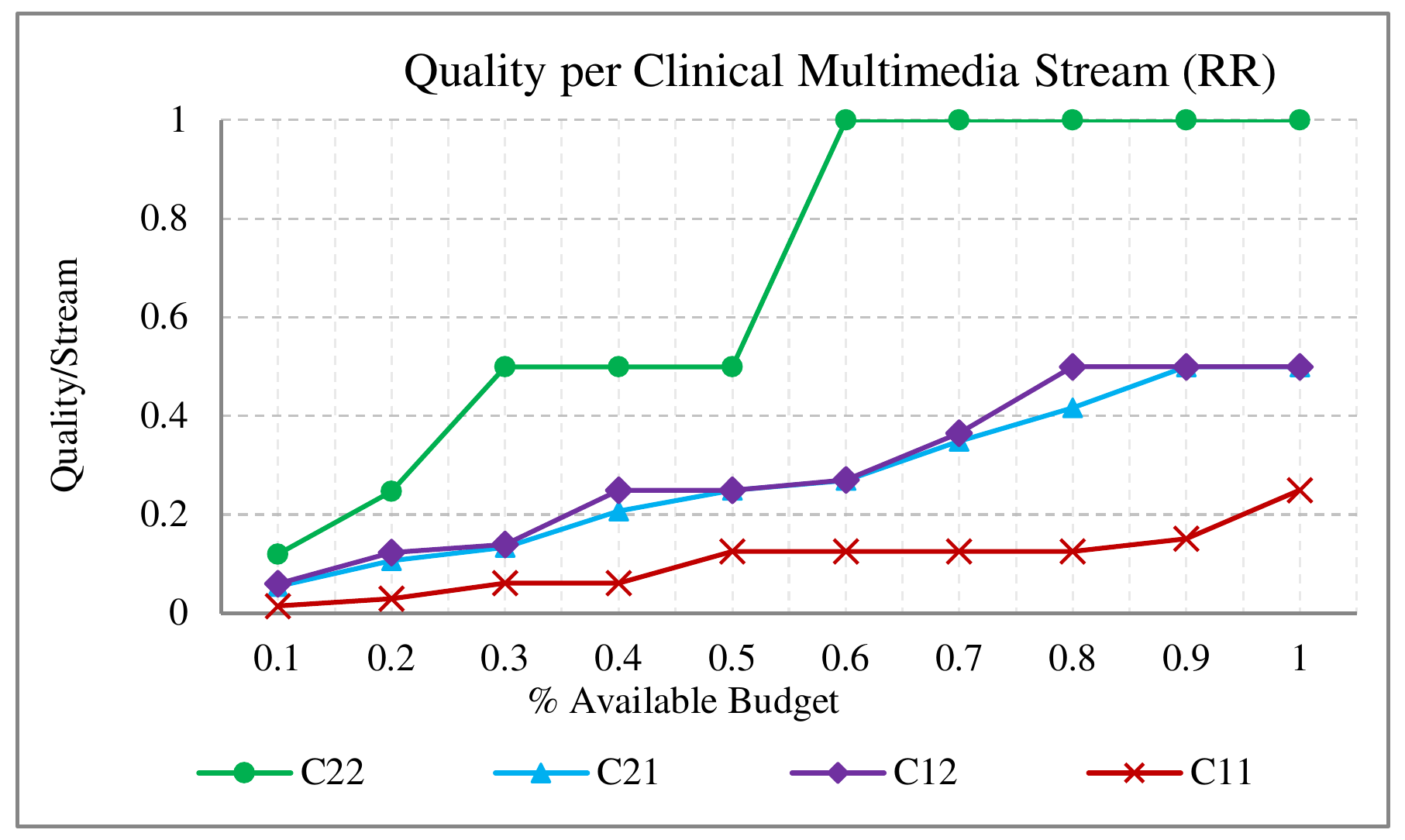}
\includegraphics[totalheight=0.21\textheight, trim = 45 240 45 240, clip = true]{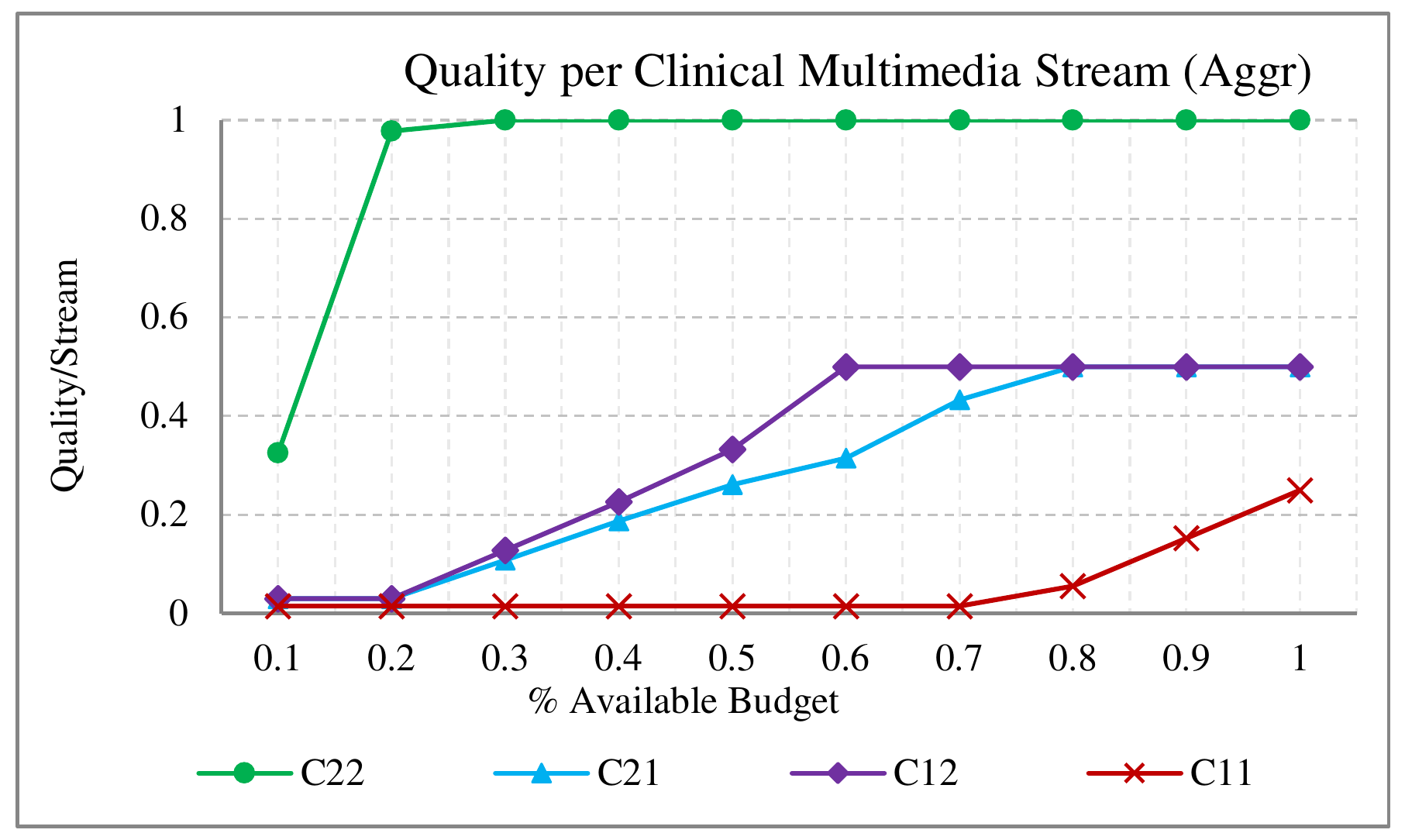}\includegraphics[totalheight=0.21\textheight, trim = 45 240 45 240, clip = true]{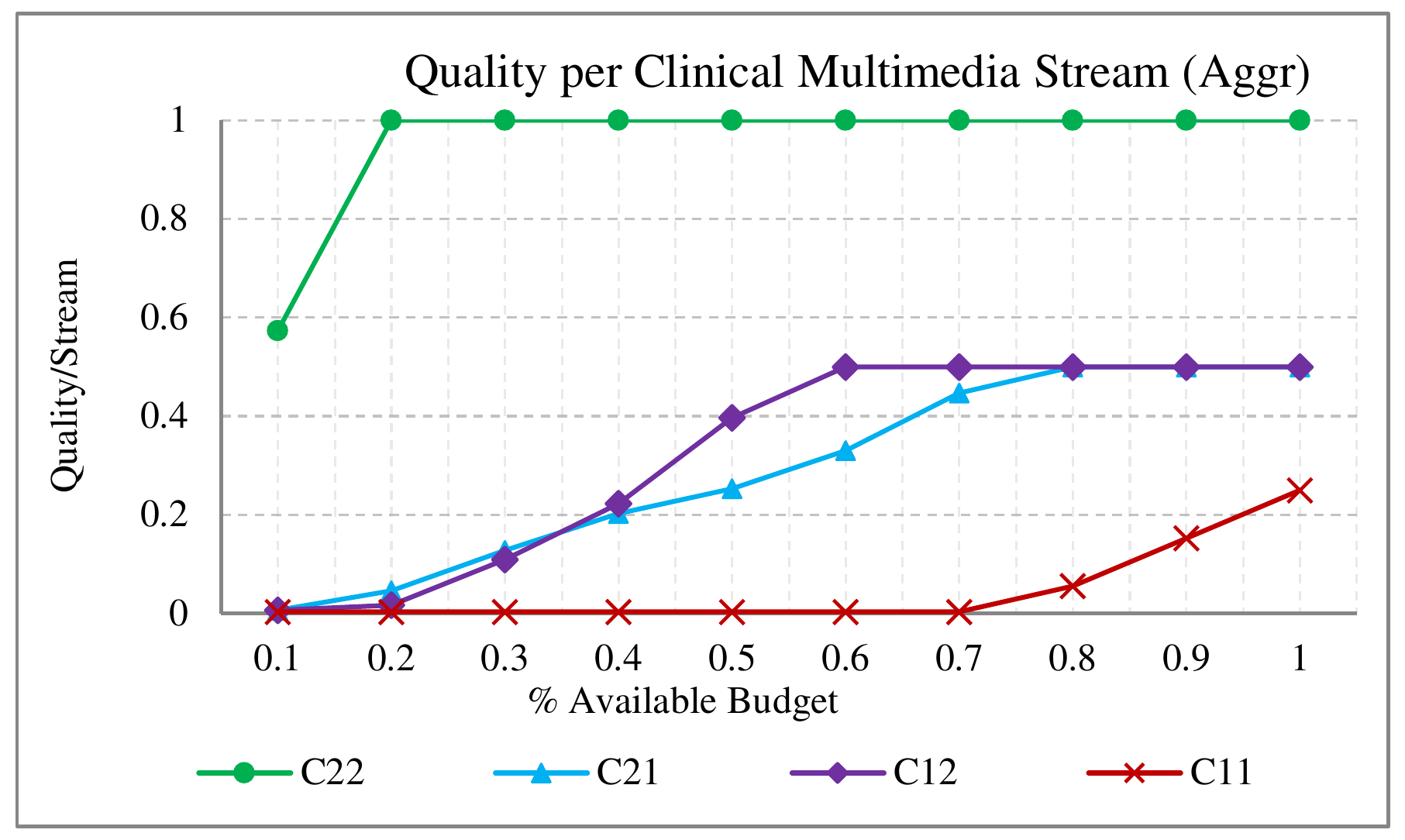}
\caption{Average quality per clinical data measured for four priority classes, and two values of $R_{max}=1/c^k$. (Top-Left) Compromise, $k=4$, (Top-Right) Compromise, $k=6$, (Middle-Left) Round-Robin, $k=4$, (Middle-Right) Round-Robin, $k=6$, (Bottom-Left) Aggressive, $k=4$, (Bottom-Right) Aggressive, $k=6$.}
\label{pdf:qps}
\end{figure*}
To assess the effectiveness of the proposed methodology in real world and to facilitate a logical interpretation of the problem to be analyzed, we conducted a communication profiling experiment in a small region of hospital health system in Illinois \cite{hosseinidataset, ichi2017}. We took a scenario where a patient is transferred from a rural hospital to a regional center hospital via a high-speed ambulance, and profiled geo-communication information including available bandwidth covering major routes from Hoopeston to Urbana as proof of concept. Hoopeston Health Center (rural hospital) is an integrated part of Carle Foundation Hospital in Urbana (center hospital), which includes medical clinic based in Hoopeston, Illinois, with multiple additional clinics serving its surrounding rural communities.

To measure the available bandwidth, we developed a mobile bandwidth profiler application in collaboration with Carle Ambulance Service to validate both the \textit{variability} and \textit{limitation} of bandwidth under our bandwidth-compliant prioritized adaptation system. We used Android SDK 25 for development, and used Google Nexus 5 smartphone as our profiling platform mounted with 4G LTE ICCID SIM Cards under both Sprint and AT\&{T} cellular networks.

Our profiler periodically samples and logs useful geo-communication information once every 4 seconds
, including: a) rate of available bandwidth, b) timestamp, c) GPS longitude, d) GPS latitude, e) altitude, and f) velocity. To implement the profiling process of bandwidth rates, the profiler client first establishes multiple TCP connections with our server over port 8080, and continuously retrieves file chunks of 1 megabits for a course of 4 seconds. Our server is a local HTTP server that we specified to minimize latency and jitter due to congestion and communication errors. As the chunks are received by the profiler, the profiler requests more file chunks throughout the fixed duration. The total size of the buffered transfers is then received, and the download speed is calculated in kbps (1 byte = 0.0078125 kilobits) given the fixed specified duration. The sampling process ends once the configured amount of duration has been reached. 

Figure \ref{profile} demonstrates only a single sample of our profiled data, the communication bandwidth under Sprint cellular network accounting for one of the major routes for a total distance of 53 miles and a total travel time of 47 minutes. The vertical axis shows the available bandwidth while the horizontal axis shows the timestamp with each point of data accounting for the four seconds of sampling period. The figure showcases interesting results to support our assumption of bandwidth \textit{variability} and \textit{limitation}. As can be clearly observed, the results show lower communication bandwidth on rural areas while they show higher bandwidth as we get closer to the urban area. It can be concluded from the results that the communication bandwidth along this route can range from as low as a few Kbps to as high as several Mbps, with most part of the route suffering from very poor communication coverage. The low communication bandwidth severely limits the amount of clinical multimedia data that can be communicated during emergency rural patient transport via an ambulance. It should be noted that the average travel speed during the profiling process was 72 miles per hour to comply with the speed limits. It is expected that the higher travel speed of an ambulance during an emergency situation further limits the available communication bandwidth.

To evaluate our proposed adaptation algorithms, we ran 10 experimental tests of simulated clinical models derived from automata of 30 different physiological states and a maximum of 15 types of various clinical multimedia data required as per each clinical state. We used NIH PhysioNet databases of recorded physiological information needed for clinical multimedia data \cite{physionet}. We classified the criticality of both the states and clinical multimedia data and determined the four priority classes $C_{11}$, $C_{12}$, $C_{21}$, and $C_{22}$ according to our definitions in Section \ref{sec:methodology}. We calculated performance measurements of our adaptation heuristics, and collected statistical including the mean and standard deviation of the results. Our performance measure is the total quality of clinical multimedia data according to Section \ref{sec:methodology}, which is a measure of the effectiveness of an approach to maximizing the total of our defined quality of clinical multimedia data based on prioritization, with larger values being more effective. In compliance with both the \textit{variability} and \textit{limitations} of the results of our bandwidth profiling, we ran our experiments with the available bandwidth $W$ set to be different percentages of \textit{S} (total of full bandwidth requirements of all clinical multimedia data). In particular, we set $W$ to 0.1\textit{S}, 0.2\textit{S}, ..., 1.0\textit{S} corresponding to 10\%, 20\%, ..., 100\% of the total bandwidth of all clinical multimedia streams. We set $R_{max}=\frac{1}{c^k}$ as the maximum possible scaling factor, in which we chose $c=2$ based on a trade-off between the different choices of sampling frequencies and a decreasing fine-grained gaps between two consequent rates, and ran our experiments for two different values of $R_{max}$. In practice, values of $R_{max}$ is prescribed by the physicians at the regional center hospital based on patient's various physiological states. We also measured the contribution of each of the four priority classes $C_{11}$, $C_{12}$, $C_{21}$, and $C_{22}$ to the total quality of clinical multimedia data for all three algorithms. 

Figure \ref{pdf:quality} shows the experimental results for normalized total quality of clinical multimedia data achieved using our three adaptation algorithms, and two different values of $R_{max}=1/c^k$ ($k=4$ and $k=6$). We used two different tuples ($p_0$, $p_1$, $p_2$), (1,2,2) and (1,3,3), with $p_1$ and $p_2$ showing first-level and second-level priorities respectively relative to $p_0$ to differentiate the priorities of clinical multimedia data in different classes. We set $p_0=1$ to normalize the priorities. As can be seen, the total effectiveness achieved by all three algorithms increases as the bandwidth increases simply because sampling frequency contributes to bandwidth which is a component of our definition of quality of clinical multimedia data. As the bandwidth budget increases, there is more space for higher sampling frequencies of clinical multimedia data, and therefore higher clinical bandwidth. Also, the total quality of clinical multimedia data increases as the ratio of $\frac{p_1}{p_0}$ or $\frac{p_2}{p_0}$ increases, confirming that our proposed criticality-aware adaptations noticeably distinguish the more critical clinical multimedia data from the less critical clinical multimedia data. As can be seen, while the figure shows a larger gap between the red line (no priority) and the green line (Prioritized (1,2,2)), the corresponding gap between the green line and the blue line (Prioritized (1,3,3)) is less. This signifies the fact that as the ratio $\frac{p_1}{p_0}$ increases towards any values higher than 3, the green and blue line will be less distinguishable, and will most likely merge in the overall trend of higher $\frac{p_1}{p_0}$ ratios. As can be witnessed already, in the range of 50\% to 70\% of the bandwidth percentile, the green line and the blue line demonstrate very close values, almost identical at 60\%. This fact shows that ratios between the smallest and largest priorities that are larger than 1:3 are not likely to be more effective. 

Figure \ref{pdf:adaptations} shows the results for average bandwidth reduction in terms of how much the sampling frequency adaptations scaled down the bandwidth, given the equation
\begin{equation*}
\scriptsize{Adaptation~Ratio}= \dfrac{\text{Total quality of clinical multimedia data \textit{after} adaptation}}{\text{Total quality of clinical multimedia data \textit{before} adaptation}}
\end{equation*}

measured for four different pairs indicating each clinical priority classes, and two different values of $R_{max}=1/c^k$ ($k=4$ and $k=6$). 
As can be seen, the higher $R_{max}$ results in more bandwidth reduction for clinical multimedia data in $C_{11}$ (corresponding to the less critical class), while preserving more of the full-bandwidth clinical multimedia streams in $C_{22}$ (corresponding to the more critical class). Clearly the larger values of $R_{max}$ work better for situations with lower communication bandwidth. Also, similar to the total quality of clinical multimedia data, these figures suggest two points. Firstly, the adaptation ratio increases as the available bandwidth increases. Secondly, the diagram of $C_{22}$ shows larger adaptation ratio compared to $C_{11}$, confirming that our proposed adaptations noticeably distinguish the more critical clinical multimedia data from the less critical clinical multimedia data, with \textit{Aggressive} and \textit{Round-Robin} causing the most, and the least differentiation, respectively, as compared to a compromise and fair differentiation resulted by \textit{Compromise}.

Figure \ref{pdf:qps} shows normalized average quality of clinical multimedia data per each clinical multimedia data for the same set of experiments. 
The diagrams confirm the conclusions derived from Figure \ref{pdf:quality} and Figure \ref{pdf:adaptations}. As can be seen, the diagram of $C_{22}$ shows larger values compared to the other priority classes. It is worth mentioning that our adaptations carry maximum effectiveness when all the clinical multimedia data in $C_{11}$ are adapted by $R_{max}$, and none of the clinical multimedia data in $C_{22}$ are adapted. This specific point is considered as the peak of quality of clinical multimedia data. As our approach adapts the clinical multimedia data in $C_{22}$, the gain in quality of clinical multimedia data brought by our approach is being decreased.

Overall, although our prioritized bandwidth-compliant clinical multimedia data adaptations do provide degradation in clinical multimedia data sampling frequencies in general, considering the communication bandwidth constraints in emergency rural ambulance transport, it is reasonable to believe that the medical community would make this small sacrifice in less-critical clinical multimedia data in exchange for respecting higher bandwidth for more critical clinical multimedia data using our adaptations system.

\section{Conclusion and Future Work}
\label{sec:conclusion}
Use of telecommunication technologies can enhance effectiveness and safety of emergency ambulance transport of a patient from rural areas to a regional center hospital. It enables remote monitoring of the patient by the physicians at the center hospital which provides vital assistance to the EMT to associate best treatments. However, the communication along the roads in rural areas range from 4G to 2G to low speed satellite links. This variable and limited communication bandwidth together with the produced mass of clinical multimedia data pose a major challenge in real-time supervision of patients.

In this paper, we take initial steps towards physiology-aware DASH, and propose bandwidth-compliant prioritized adaptation techniques to manage transmission of massive clinical multimedia data during ambulance transport with limited bandwidths. We use the concepts of model-driven clinical automata, and exploit the semantics relation of limited communication bandwidth with criticality of clinical multimedia data in a physiology-aware manner. In collaboration with Carle Foundation Hospital, we developed a profiler, and profiled the communication bandwidth for a realistic emergency rural ambulance transport to support our experiments. Our initial evaluation results show that our adaptations can improve the effectiveness of emergency care by transferring the most critical clinical information given the limited bandwidth.

In the future, we plan to derive a more sophisticated measures of quality of clinical multimedia data and integrate our objective function with subjective clinical metrics. We also plan to add mathematical analysis for approximation bounds. In the near future, we will clinically validate our system in a real emergency rural ambulance transport.
\bibliographystyle{IEEEtran}
\bibliography{sigproc-Revised}

\begin{IEEEbiography}[{\includegraphics[width=1in,height=1.25in,clip,keepaspectratio]{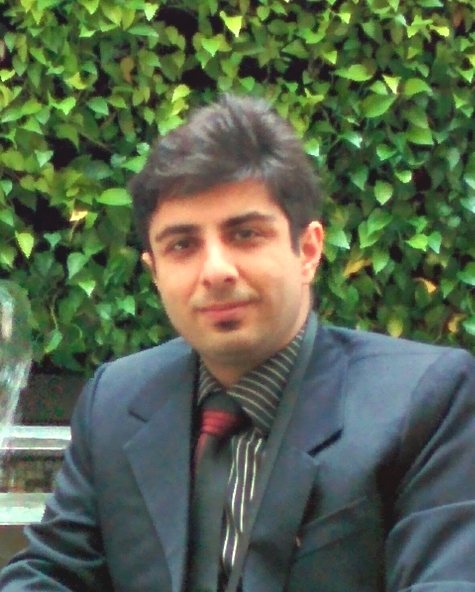}}]{Mohammad Hosseini}
received the BSc degree from Sharif University of Technology in 2011, the MSc degree from Simon Fraser University in 2013, and the Ph.D. degree in Computer Science from University of Illinois at Urbana-Champaign (UIUC) in 2017. His research background mainly lies in multimedia systems, especially optimization in mobile multimedia, 3D technologies, and VR. He has served as TPC member, reviewer, and guest reviewer at major workshops, conferences, and journals including ACM MM, ACM MMSys, IEEE TMM, IEEE ICME, ACM MoVid, IEEE Netgames, Springer MMSJ, etc. Dr. Hosseini has internship experiences at multiple research labs including Qualcomm Research, Fujitsu Labs of America, and Adobe Research, where he was recognized with several intern awards.
\end{IEEEbiography}

\begin{IEEEbiography}[{\includegraphics[width=1in,height=1.25in,clip,keepaspectratio]{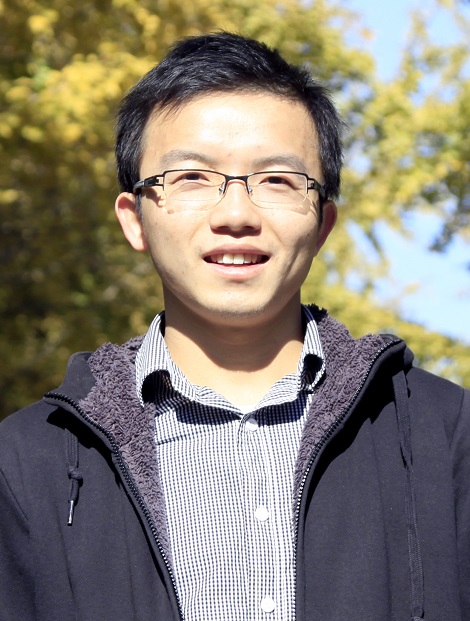}}]{Yu Jiang}
received the BS degree in software engineering from Beijing University of post and telecommunication, in 2010, and the PhD degree in Computer Science from Tsinghua University, in 2015. He worked as a Postdoc researcher in the department of computer science of University of Illinois at Urbana-Champaign, in 2016.

Dr. Jiang is now an assistant professor in the School of Software, Tsinghua University. 
His current research interests include program analysis, formal computation model, formal verification and their applications in embedded systems, safety analysis and assurance of cyber-physical system.
\end{IEEEbiography}

\begin{IEEEbiography}[{\includegraphics[width=1in,height=1.25in,clip,keepaspectratio]{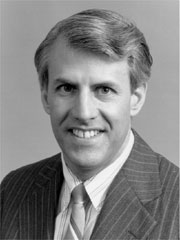}}]{Richard R. Berlin}
Richard Berlin received his MD from SUNY-Downstate in 1974 and MBA from CUNY-Baruch in 1992.​
Dr. Richard Berlin is a Level 1 trauma surgeon at Carle Foundation Hospital, and an adjunct associate professor in the Department of Computer Science at UIUC.  He  served as the medical director for health systems at Health Alliance, the regional Health Maintenance Organization (HMO) for central Illinois.  The co-author of the text Healthcare Informatics (Springer), Dr. Berlin's interests include Systems Medicine summarized in a most recent (Springer journal) publication, Systems Medicine - Complexity Within, Simplicity Without.
\end{IEEEbiography}

\begin{IEEEbiography}[{\includegraphics[width=1in,height=1.25in,clip,keepaspectratio]{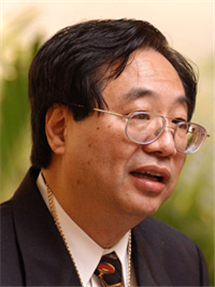}}]{Lui Sha}
received the Ph.D. degree from Carnegie Mellon University, Pittsburgh, PA, in 1985. He is currently a Donald B. Gillies Chair Professor of computer science at the University of Illinois at Urbana Champaign. His work on real-time computing is supported by most of the open standards in real-time computing and has been cited as a key element to the success of many national high-technology projects including GPS upgrade, the Mars Pathfinder, and the International Space Station. Professor Sha is a recipient of the prestigious 2016 IEEE Simon Ramo Medal for technical leadership and contributions to fundamental theory, practice, and standardization for engineering real-time systems.
\end{IEEEbiography}

\begin{IEEEbiography}[{\includegraphics[width=1in,height=1.25in,clip,keepaspectratio]{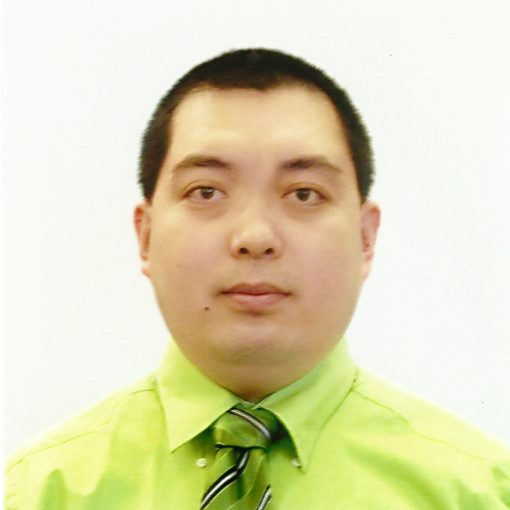}}]{Houbing Song}
received the Ph.D. degree in electrical engineering from the University of Virginia, Charlottesville, VA, in August 2012.

In August 2017, he joined the Department of Electrical, Computer, Software, and Systems Engineering, Embry-Riddle Aeronautical University, Daytona Beach, FL, where he is currently an Assistant Professor and the Director of the Security and Optimization for Networked Globe Laboratory (SONG Lab, www.SONGLab.us). He was a faculty member of West Virginia University from August 2012 to August 2017. He has served as an Associate Technical Editor for IEEE Communications Magazine since 2017. He is the editor of 4 books and the author of more than 100 articles. His research interests include cyber-physical systems, internet of things, cloud computing, big data analytics, connected vehicle, wireless communications and networking, and optical communications and networking. Dr. Song is a senior member of the ACM.

\end{IEEEbiography}

\end{document}